\documentclass[letterpaper,twocolumn,prb,aps,superscriptaddress,amsmath,amssymb,floatfix]{revtex4-2}
\usepackage{mathptmx}
\usepackage[utf8]{inputenc}
\usepackage{color}
\usepackage{amsmath}
\usepackage{mathtools}
\usepackage{amssymb}
\usepackage{graphicx}
\usepackage{esint}
\usepackage[unicode=true,
 bookmarks=true,bookmarksnumbered=false,bookmarksopen=false,
 breaklinks=false,pdfborder={0 0 1},backref=false,colorlinks=true,colorlinks=true,citecolor=blue,linkcolor=blue,urlcolor=blue]
 {hyperref}
\hypersetup{
 linkcolor=black,urlcolor=black,citecolor=black,pdfstartview={FitH},hyperfootnotes=false}

\makeatletter

\usepackage{textcomp}
\usepackage{epstopdf}

\pdfpageheight\paperheight
\pdfpagewidth\paperwidth

\@ifundefined{textcolor}{}{%
 \definecolor{BLACK}{gray}{0}
 \definecolor{WHITE}{gray}{1}
 \definecolor{RED}{rgb}{1,0,0}
 \definecolor{GREEN}{rgb}{0,1,0}
 \definecolor{BLUE}{rgb}{0,0,1}
 \definecolor{CYAN}{cmyk}{1,0,0,0}
 \definecolor{MAGENTA}{cmyk}{0,1,0,0}
 \definecolor{YELLOW}{cmyk}{0,0,1,0}
}

\usepackage{xcolor}\usepackage{soul}
\definecolor{blue}{rgb}{0,0,1}
\definecolor{red}{rgb}{1,0,0}
\definecolor{green}{rgb}{0,1,0}

\usepackage{soul}

\usepackage{graphicx}
\usepackage{array}
\usepackage{multirow,booktabs}
\usepackage{bm}
\usepackage[T1]{fontenc}
\usepackage{color}
\usepackage{bbm}
\begin{document}
\renewcommand{\thetable}{\arabic{table}}
\renewcommand{\tablename}{\textbf{Table}}
\title{Generalized Number-Phase Lattice Encoding of a Bosonic Mode for Quantum Error Correction}
\author{Dong-Long Hu}
\affiliation{School of Physics, Sun Yat-sen University, Guangzhou 510275, China }
\author{Weizhou Cai}
\affiliation{CAS Key Laboratory of Quantum Information, University of Science and Technology of China, Hefei 230026, China}
\author{Chang-Ling Zou}
\email{clzou321@ustc.edu.cn}
\affiliation{CAS Key Laboratory of Quantum Information, University of Science and Technology of China, Hefei 230026, China}
\affiliation{Hefei National Laboratory, Hefei 230088, China}
\author{Ze-Liang Xiang}
\email{xiangzliang@mail.sysu.edu.cn}
\affiliation{School of Physics, Sun Yat-sen University, Guangzhou 510275, China }
\affiliation{State Key Laboratory of Optoelectronic Materials and Technologies, Sun Yat-sen University, Guangzhou 510275, China}
\date{\today}

\begin{abstract}
\textbf{ Bosonic systems offer unique advantages for quantum error correction, as a single bosonic mode provides a large Hilbert space to redundantly encode quantum information. However, previous studies have been limited to exploiting symmetries in the quadrature phase space. Here we introduce a unified framework for encoding a qubit utilizing the symmetries in the phase space of number and phase variables of a bosonic mode. The logical codewords form lattice structures in the number-phase space, resulting in rectangular, oblique, and diamond-shaped lattice codes. Notably, oblique and diamond codes exhibit a number-phase vortex effect, where number-shift errors induce discrete phase rotations as syndromes, enabling efficient correction via phase measurements. These codes show significant performance advantages over conventional quadrature codes against dephasing noise in the potential one-way quantum communication applications. Our generalized number-phase codes open up new possibilities for fault-tolerant quantum computation and extending the quantum communication range with bosonic systems.
}
\end{abstract}
\maketitle

\noindent \textbf{\large{}Introduction}{\large\par} 
\noindent Quantum error correction (QEC)~\cite{Nielsen_Chuang_2010,Barbara2015} is essential for realizing reliable quantum computation and communication in the presence of unavoidable environmental noise~\cite{Cory1998,Knill2001a,Waldherr2014,Abobeih2022,Chiaverini2004,Anderson2021,Schindler2011,Egan2021,Postler2022,Yao2012,Takeda2022,Kelly2015,Google2021,Krinner2022,Zhao2022,Google2023,Reed2012,Corcoles2015}. While quantum information is commonly encoded in discrete variables of multiple physical qubits, single-mode bosonic encodings offer a hardware-efficient alternative by leveraging the infinite-dimensional Hilbert space of a single oscillator. Moreover, bosonic information carriers play a crucial role in transferring information over distances, making bosonic encodings indispensable in quantum communications, and distributed quantum computation and sensing~\cite{cai2021}. The ability to encode and protect quantum information in bosonic modes, such as optical or microwave cavities, has attracted considerable attention in recent years. Notably, experiments with superconducting systems~\cite{Blais2021} have achieved milestones by reaching the break-even point in QEC with bosonic codes~\cite{Ofek2016,Sivak2023,Ni2023,Cai2024a}, demonstrating their immense potential for fault-tolerant quantum applications.

Existing bosonic QEC codes can be broadly categorized based on the symmetries they exploit in the phase space of a bosonic mode. The Gottesman-Kitaev-Preskill (GKP) codes~\cite{Gottesman2001}, for instance, leverage the displacement symmetry to correct random displacement errors in the quadrature variables. On the other hand, codes like the cat and binomial codes utilize the rotational symmetry in phase space~\cite{Cochrane1999,Leghtas2013,Michael2016,Grimsmo2020}, which corresponds to a certain generalized parity of the photon number distribution. These codes are designed to tolerate photon loss or gain errors, while the code distance is determined by the photon-number parity of the codewords. However, beyond the conventional quadrature variables, the photon number and phase can also be treated as canonical variables for a bosonic mode. The potential symmetries and encoding schemes in this generalized number-phase (NP) space have not been fully explored.

In this paper, we propose a unified framework for constructing bosonic QEC codes, which embody a lattice structure in the NP space, encompassing the known cat and binomial codes as rectangular lattices. Remarkably, we uncover oblique and diamond lattice codes that exhibit a novel NP vortex effect: number-shift errors in these codes induce discrete phase rotations as error syndromes. This unique feature enables us to identify and correct errors by measuring only the phase variable of the bosonic mode. We demonstrate that our generalized codes, particularly the oblique and diamond lattice codes, exhibit enhanced performance in the presence of practically relevant noise channels that inflict both amplitude damping and dephasing. Furthermore, we propose a hardware-efficient quantum error correction protocol that is compatible with one-way quantum communication schemes. Our framework not only unifies the existing bosonic codes but also paves the way for discovering new codes with desirable properties. The NP lattice codes introduced here have the potential to significantly advance fault-tolerant quantum computation and extend the range of quantum communication~\cite{Briegel1998,Jiang2009,Lo2014,Takeoka2014,li2017,Rozpedek2021,Wu2022,li2024}.

\begin{figure*}[t]	
\centering
\includegraphics[width=0.95\linewidth]{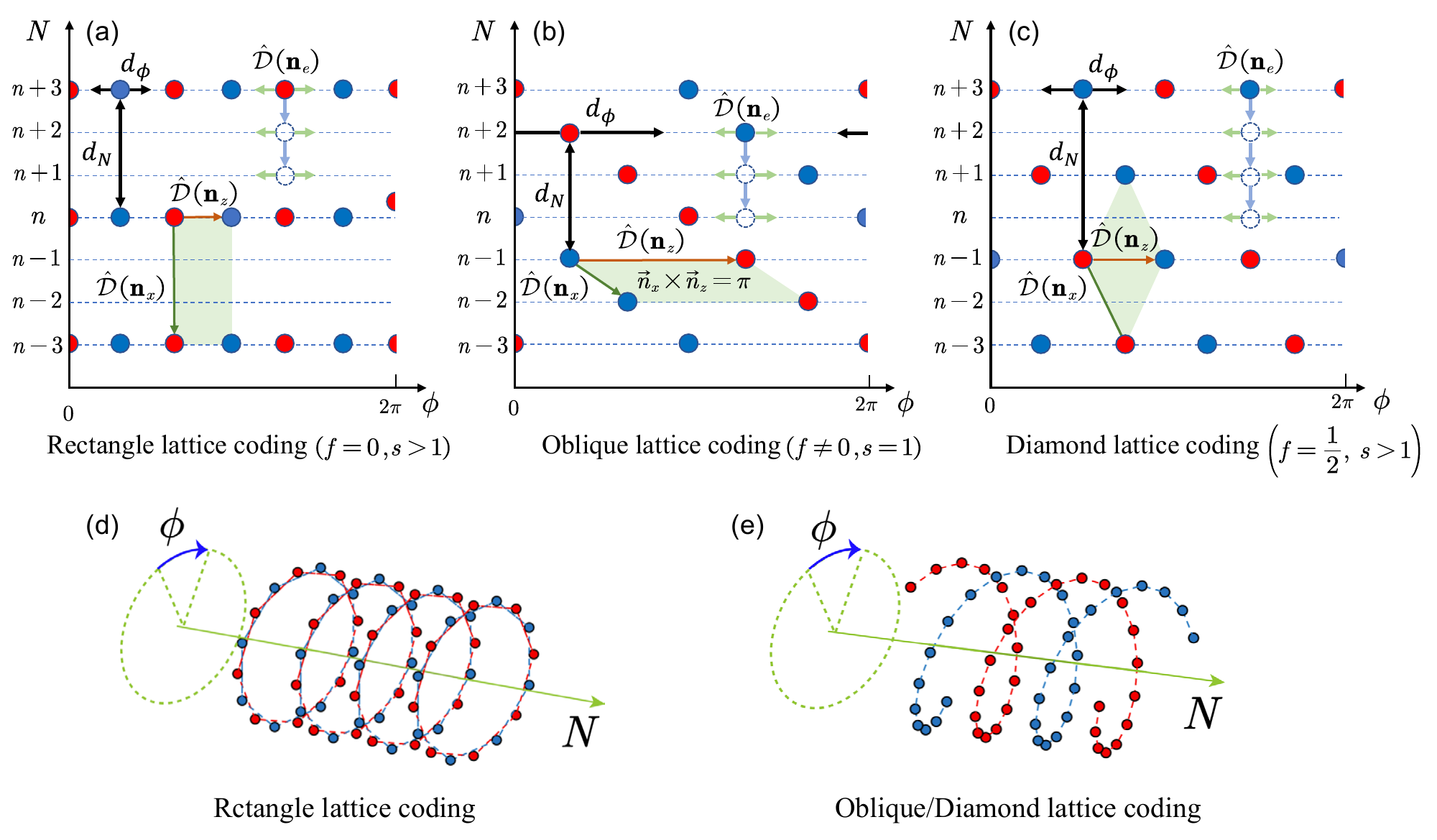}
\caption{Schematics of three types of generalized NP codes in phase space defined by canonical number and phase variables. (a)-(c) show the generalized NP codes with rectangle, oblique, and diamond lattice coding. Red and blue circles represent probability peaks of NP Wigner function of dual logical states $|+\rangle_L$ and $|-\rangle_L$, respectively. $\hat{\mathcal{D}}(\mathbf{n}_{x})$ and $\hat{\mathcal{D}}(\mathbf{n}_{z})$ are the logical $\bar{X}$ and $\bar{Z}$ operations, which are represented as deep green and orange arrows, respectively.  $\hat{\mathcal{D}}(\mathbf{n}_{e})$ is the NP-shift error, represented by light green and light blue arrows. (d) and (e) are schematic diagrams of the generalized NP codes in NP space, which depict the codes rolling along the phase direction to form a cylindrical surface.}
\label{fig1}
\end{figure*}

\smallskip{}

\noindent \textbf{\large{}Results}{\large\par}

\smallskip{}

\noindent \textbf{Generalized number-phase encoding}

\noindent Inspired by the construction of GKP codes in the quadrature phase space, we introduce generalized displacement operators in the NP space. Consider a bosonic mode described by the annihilation and creation operators $\hat{a}$ and $\hat{a}^\dagger$, respectively,  $\hat{n}=\hat{a}^\dagger \hat{a}$ represents the excitation-number operator. The NP displacement operator reads
\begin{align}
   \hat{\mathcal{D}}(\mathbf{n})\equiv\exp(\frac{il\phi}{2})\hat{R}(\phi)\hat{\Sigma}_l.
    \label{eq1}
\end{align}
Here, $\hat{R}(\phi)=\exp(i\hat{n}\phi)$ is the phase rotation operator, $\mathbf{n}=(l,\phi)$ is a two-dimensional vector with the limitation $l\in\mathbb{Z}$ and $\phi\in\mathbb{R}$. $\hat{\Sigma}_l\equiv\sum_{n}|n\rangle \langle n+l|$ is the Fock ladder shift operator, which connects to the annihilation operator with the relation $\hat{a}=\sqrt{\hat{n}+1}\hat{\Sigma}_1$. Note that $\hat{\Sigma}_1^\dagger=\hat{e^{-i\phi}}$ is the Susskind-Glogower exponential phase operator~\cite{Susskind1964}, which is well-known to be nonunitary. Nonetheless, the nonunitary does not prevent the NP displacement operators from being a set of complete operator basis, because they satisfy the same commutation relation
\begin{align}
    \hat{\mathcal{D}}(\mathbf{n})\hat{\mathcal{D}}(\mathbf{n}')=e^{i\mathbf{n}\times\mathbf{n}'}\hat{\mathcal{D}}(\mathbf{n}')\hat{\mathcal{D}}(\mathbf{n}),
    \label{eq2}
\end{align}
and the orthogonality relation $\mathrm{Tr}(\hat{\mathcal{D}}^\dagger(\mathbf{n}')\hat{\mathcal{D}}(\mathbf{n}))=2\pi\delta^2(\mathbf{n}-\mathbf{n}')$ with the conventional displacement operators. In this sense, an arbitrary operator satisfying $\mathrm{Tr}(\hat{E}^\dagger E)<\infty$ can be expanded by using the NP displacement operator basis, 
\begin{align}
   \hat{E}=\int \frac{d\mathbf{n}}{2\pi}~\mathrm{Tr}(\hat{\mathcal{D}}^\dagger(\mathbf{n})\hat{E})\hat{\mathcal{D}}(\mathbf{n}).
\end{align}
That is, a noise operator $\hat{E}$ for bosonic modes can be represented as a series of NP-displacements $\hat{\mathcal{D}}(\mathbf{n})$ with the weight $\mathrm{Tr}(\hat{\mathcal{D}}^\dagger(\mathbf{n})\hat{E})$. 

In general, a logical qudit with a finite dimension $d$ can be encoded in a bosonic mode, which provides an infinite-dimensional Hilbert space to manage errors, thereby resulting in error-correcting bosonic codes. Here we focus on the simplest case, encoding a qubit into a bosonic mode for removing small NP displacements errors $\hat{\mathcal{D}}(\mathbf{n}_e)$. This approach is referred to as the generalized NP encoding of a bosonic mode, with the codewords $|0\rangle_L$ and $|1\rangle_L$ representing two highly symmetric superposition states of NP variables. To provide a more intuitive representation of codewords in NP phase space, recall that $\mathcal{W}_{\rho}(\mathbf{c})$ represents the NP Wigner function of the quantum state $\rho$. Here, $\mathbf{c}=(n,p)$ specifies arbitrary coordinates in two-dimensional NP phase space, subject to the constraints $n\in\mathbb{N}$ and $p\in[0,2\pi)$. The detailed definition of the NP Wigner function can be found in the supplementary Note 1, including Ref.~\cite{Vaccaro1995}.

Specifically, the codewords design of such an encoding is based on Eq.~(\ref{eq2}), the commutation between two NP displacement operators results solely in a phase factor $e^{i\mathbf{n}\times\mathbf{n}'}$. Here, $|\mathbf{n}\times\mathbf{n}'|$ should be understood as the area of a parallelogram in NP space, with $\mathbf{n}$ and $\mathbf{n}'$ as its defining edges. Notably, the two NP displacement operators $ \hat{\mathcal{D}}(\mathbf{n}_x)$ and $\hat{\mathcal{D}}(\mathbf{n}_z)$ correspond to the logical Pauli operators, satisfying the anti-commutation relation $\bar{X}\bar{Z}=-\bar{Z}\bar{X}$ when the condition 
\begin{align}
|\mathbf{n}_x\times\mathbf{n}_z|=\pi 
\label{eq4}
\end{align}
is satisfied. Consequently, they should be square to two stabilizers $\hat{S}_x\equiv \hat{\mathcal{D}}(2\mathbf{n}_x)$ and $\hat{S}_z\equiv\hat{\mathcal{D}}(2\mathbf{n}_z)$, thus the codewords of generalized NP codes are defined by the simultaneous +1 eigenstates of the two stabilizers. In other words, the codespaces $\bar{P}_{\mathrm{code}}=|0\rangle\langle 0|_L+|1\rangle\langle 1|_L$ of generalized NP codes have discrete translational invariance in NP variables, forming as a lattice with the cell area $\pi$ in NP space. 

Since any pair of $\mathbf{n}_x$, $\mathbf{n}_z$ fulfilling Eq.~(\ref{eq4}) is valid, there are infinite types of generalized NP codes. Fortunately, due to the $2\pi$ periodicity of the phase variable, any pair of $\mathbf{n}_x,\mathbf{n}_z$ satisfying Eq.~(\ref{eq4}) can be equivalently expressed in a gauge
\begin{align}
    \mathbf{n}_x=(s,\frac{f\pi}{s}), \quad \mathbf{n}_z=(0,\frac{\pi}{s}).
    \label{eq5}
\end{align}
Here $s\geq1$ is a positive integer that defines the rotation symmetry $\hat{S}_z=\hat{R}(2\pi/s)$ of codespaces, and $|f=p/q|\in[0,1)$ is a fractional number given by two coprime integers $p\in\mathbb{Z}$ and $q\in\mathbb{Z}^+$. The set of parameters $(s,f)$ defines the lattice structure of codespaces in NP space, and the NP Wigner function of the codespaces can be directly obtained as (unnormalized)
\begin{align}
    \mathcal{W}_{\bar{P}_{\mathrm{code}}}(\mathbf{c})=\sum_{r=0,t=0}^{\infty}\delta^2(r\mathbf{n}_x^*+t\mathbf{n}_z+\mathbf{n}_0),
    \label{eq6}
\end{align}
where $\mathbf{n}_x^*\equiv(s,-f\pi/s)$, and $\mathbf{n}_0=\nu_x\mathbf{n}_x^*+\nu_z\mathbf{n}_z$ with $\nu_x , \nu_z\in[0,1)$ serves as a selectable origin in NP space. Note that the phase coordinates on the right side of Eq.~(\ref{eq6}) may exceed the defined range $[0,2\pi)$, which is insignificant and can be returned to the defined range through the $2\pi$ periodicity of the phase variable.

In Figs.~\ref{fig1} (a)-(c), we graphically illustrate the codespaces of three typical generalized NP codes in NP phase space as examples, where red and blue circles represent the peaks of the NP Wigner function of codewords $|+\rangle_L$ and $|-\rangle_L$, respectively. As expected, the codespaces of generalized NP codes are lattices in NP phase space, which can be regarded as the rectangle (R-NP), oblique (O-NP), and diamond NP (D-NP) codes determined by the shape of the lattice cell. Now, let us imagine the two-dimensional NP phase space plane rolled along the phase direction into a cylindrical surface, and the NP lattice codes will present a novel 3D physical picture, as shown in Figs.~\ref{fig1}(d) and (e). The R-NP codes are a series of circular rings along the number direction with an equal distance $s$. Conversely, NP codes with $f\neq0$ are an NP vortex along the number direction due to the non-trivial parameter $f$, leading to a hybridization of number and phase variables.

Moreover, such a graphical representation of codespaces in NP phase space can directly give the code distance of the canonical number variable $d_{N}=qs$ ($p\neq 0$) and the canonical phase variable $d_{\phi}=\pi/s$. Thus we can define the correctable error set $\bar{E}$ of the generalized NP codes for exactly satisfying the Knill-Laflamme (KL) conditions $\langle\mu|\hat{E}^\dagger_j\hat{E_k}|\nu\rangle_L=H_{j,k}\delta_{\mu,\nu}$, where $H$ is a Hermitian matrix, and $\mu$ and $\nu$ run the logical $0$ and $1$~\cite{Knill1997}. If one concentrates to remove the number-shift errors caused by excitation loss or gain, the correctable error set can be selected as $\bar{E}=\{\hat{\mathcal{D}}(\mathbf{n}_e)\}$ with the vector $\mathbf{n}_e$ in the region
\begin{align}
    \mathbf{n}_e\in\{~(l_e,\phi_{e})~|-G\leq l_e\leq L,~~|\phi_{e}|<\frac{\pi}{2d_N} \},
    \label{eq7}
\end{align}
where $G$ and $L$ are two positive integer which satisfies $G+L=d_N-1$. Particularly, another useful correctable error set with $\mathbf{n}_e\in\{~(0,\phi_{e})~|~|\phi_{e}|<{d_{\phi}}/{2} \}$ focus on removing the pure rotation errors caused by quantum dephasing, which sacrifices a portion of the ability to correct number-shift errors as a cost. The error syndromes for $\hat{\mathcal{D}}(\mathbf{n}_e)$ need to be identified through the measurement of the generalized number parity $k\in[0,s-1]$ and the rotated phase angle $\bar{\phi}=mf\pi/s+\phi_e$ of the noisy codespaces, thereby different $\mathbf{n}_e=(l_e,\phi_e)$ can be distinguished via the relation $l_e=sm+k$ and $\phi_e=\bar{\phi}-mf\pi/s$ with $m\in[\lfloor L/s\rfloor-q+1,\lfloor L/s\rfloor]$, and $\lfloor\cdot\rfloor$ is the floor function. The syndromes of these NP displacement errors for generalized NP codes with different structures are summarized in Table~\ref{table1}. The experimental realization of the error syndrome measurement will be discussed in the following.   
\begin{table}[thb]
\centering
\caption{Examples and comparisons of generalized NP codes. $s,~f,~\{\theta_n\}$ are the parameters of generalized NP codes, which are defined in Eq.~\eqref{eq8}. The first and second numbers of real array $(k,\bar{\phi})$ represent error syndromes read out by number parity measurement and phase measurement, respectively.}
\renewcommand{\arraystretch}{2}
\begin{tabular}{m{1cm}<{\centering} m{0.9cm}<{\centering} m{1cm}<{\centering} m{1.8cm}<{\centering} m{3cm}<{\centering}}
\hline\hline
Lattice & $s$ & $f=\frac{p}{q}$ & $\{\theta_n\}$ & syndromes for $\hat{D}(\mathbf{n}_e)$ \\\hline\hline
Rect.  & $>1$ & $0$ & $\sqrt{2^{-K}\binom{K}{n}}$ & $(k,\phi_e)$ \\\hline
Obl.  & $1$ & $\neq 0$ & $\langle n|\alpha,r\rangle$ & $(0,mf\pi+\phi_e)$\\\hline
Diam. & $>1$ & $\frac{1}{2}$ & $\langle n|\alpha,r\rangle$ & $(k,\frac{mf\pi}{s}+\phi_e)$ \\\hline\hline
\end{tabular}
\label{table1}
\end{table}

In experiments, the ideal generalized NP codes introduced above are unavailable. Therefore, it is necessary to encode a normalized series $\{\theta_n\}$ as the amplitude of the Fock basis $|sn\rangle$ to restrict the code state energy, where the normalization of $\theta_n$ is $\sum_n|\theta_n|^2=1$. Then, the qubit states of generalized NP codes can be defined in a logical $X$ basis by using the Fock states as
\begin{align}
    |\pm\rangle_L(s,f,\theta_n)=\exp(-\frac{if\pi\hat{n}^2}{2s^2})\sum_{n=0}^{\infty}(\pm)^{ n}\theta_{n}|sn\rangle,
    \label{eq8}
\end{align}
and the qubit states in logical $Z$ basis are $|\mu\rangle_L=(|+\rangle+(-)^{\mu}|-\rangle_L)/\sqrt{2}$, where $\mu=0,1$ and the mean excitation of such code states is restricted as $\bar{n}_{\mathrm{code}}=s\sum_n n|\theta_n|^2$. Under encoding of the series $\{\theta_n\}$, the generalized NP codes with finite energy still have the perfect discrete phase rotation invariance, but the discrete translation invariance in $\mathbf{n}_x$ direction is approximately satisfied. In comparison to the parameters $s$ and $f$, which primarily govern the ability against the excitation loss, the Fock amplitude $\theta_n$ dictates the capacity to resist the pure phase rotation. For R-NP codes, $\{\theta_n\}$ can give the Helovo's phase uncertainty~\cite{Holevo2011} $\Delta_{\phi}=|\sum_n\theta_n\theta_{n+1}|^{-2}-1$ of the codeword $|\pm\rangle_L$. A smaller phase uncertainty indicates that the rotated codeword can be distinguished better by using the canonical phase measurement~\cite{Berni2015,Daryanoosh2018,Holevo2006,Martin2020}. The positive operator-valued measure (POVM) element of the canonical phase measurement is $\hat{\mathcal{M}}_X=(2\pi)^{-1}\sum_{n,m}e^{i(n-m)X}|n\rangle\langle m|$, and the associated identity relation is $\int_{0}^{2\pi}dX\hat{\mathcal{M}}_X=\hat{I}$. Generally,  $\Delta_{\phi}$ is the uncertainty of the codeword in the direction orthogonal to $\mathbf{n}_x$ in NP space. On the other hand, since $|\sum_n\theta_n\theta_{n+1}|=|\langle\mu_L|\hat{\mathcal{D}}(\mathbf{n}_x)|\mu_L\rangle|$ that measures the NP-version translation invariance, it also can be understood as a metric to quantify how close the code states are to the ideal NP codes. Detailed discussion on the relation between $\{\theta_n\}$ and phase uncertainty is included in the supplementary Note 2.  

Though the valid choice of $\{\theta_n\}$ is endless, here we concentrate a class of Fock-amplitude that is $\theta_n=\langle n|\alpha,r\rangle$, and $|\alpha,r\rangle=\hat{D}(\alpha)\hat{S}(r)|0\rangle$ is a pure Gaussian state, and $\hat{S}(r)=\exp[\frac{1}{2}(r^*\hat{a}^2-r(\hat{a}^\dagger)^2)]$ is the original squeezing operator~\cite{Weedbrook2012}. Unlike cat codes, NP codes with such a Fock amplitude allow the KL cost function to be exponentially suppressed as $\alpha^2$ increases, which is not related to the parameters $s$ and $f$, that is, there does not exist a sweet spot like cat codes\cite{Albert2018}. In addition, the parameters $\alpha$ and $r$ of NP codes can be optimized to obtain a maximal QEC performance for a given noise channel.

The generalized NP codes defined in Eq.~(\ref{eq8}) also contain many well-known bosonic error-correcting codes. For example, the rotation-symmetric codes~\cite{Grimsmo2020}, including the binomial and cat codes, are special cases of R-NP codes with different Fock amplitudes $\{\theta_n\}$. Moreover, the phase-engineered code has been proposed to further improve the QEC performance of R-NP codes, which actually is a special case of D-NP codes~\cite{Li2021}. Lastly, although NP codes and GKP codes both admit a lattice description in the phase space of a bosonic mode, these two codes are non-trivial with respect to each other. This is because a finite NP displacement (stabilizer) cannot be obtained by superposing finite quadrature displacements, and vice versa.

From Eq.~(\ref{eq8}), the mutual conversion of two different NP lattice structures $(s,f_1)\leftrightarrow(s,f_2)$ of generalized NP codes can be achieved by the interface gate 
\begin{equation}
  \hat{U}_s(\Delta f)\coloneqq \exp(-\frac{i\Delta f\pi\hat{n}^2}{2s^2}),
  \label{eq9}
\end{equation}
and it's inverse $\hat{U}^\dagger_s (\Delta f)$, where $\Delta f=f_2-f_1$. The interface gate can change the codespace structure through modifying the Pauli X operator $\hat{\mathcal{D}}(\mathbf{n}_x)$ to be 
\begin{equation}
  \hat{U}_s(\Delta f)\hat{\mathcal{D}}(\mathbf{n}_x)\hat{U}^\dagger_s(\Delta f)=\hat{\mathcal{D}}[\mathbf{n}_x+(0,\frac{\Delta f\pi}{s})],
\end{equation}
without affecting the Pauli Z operator $\hat{\mathcal{D}}(\mathbf{n}_z)$. This is based on the relations $\hat{U}_s(\Delta f)\hat{\Sigma}_s=\hat{\mathcal{D}}(0,\Delta f\pi/s)\hat{\Sigma}_s\hat{U}_s(\Delta f)$ and $[\hat{U}_s(\Delta f),\hat{R}(\phi)]=0$. Such a modification of codespaces can change the code distance from $d_N=sq_1$ to $d_N=sq_2$, while another code distance $d_\phi$ remains unchanged. Here, $q_1$ and $q_2$ are the denominator of fractions $f_1$ and $f_2$, respectively.

Another main use of the interface gate is to act as a bridge between the R-NP codes $(s,0)$ and generalized NP codes $(s,f)$, with a shared code parameter $s$. Especially, since any phase rotation for generalized NP codes commutes with the interface gate, it can be effectively readout via the canonical phase measurement once the generalized NP codes are interfaced into R-NP codes to minimize the canonical phase uncertainty of the codeword $|\pm\rangle_L$. This is because the probability distribution of the codeword $|\pm\rangle_L$ is localized in the direction orthogonal to $\mathbf{n}_x$ in NP space, as shown in the Fig.~\ref{fig1}, and only R-NP codes satisfy the orthogonality between $\mathbf{n}_x$ and the canonical phase axis. This method is a foundation to readout the error syndromes characterized by phase rotation for the QEC of generalized NP codes we will propose next. For simplicity, we use the notation $\hat{U}_f\coloneqq\hat{U}_s(f)$ and $\hat{U}_f^\dagger$ to indicate the codespaces transformation $(s,0)\rightarrow(s,f)$ and $(s,f)\rightarrow(s,0)$, respectively. The graphical representation of the interface between generalized NP codes and R-NP codes can be found in the supplementary Note 3.
 
\begin{figure}[tbp]	
\centering
\includegraphics[width=1\linewidth]{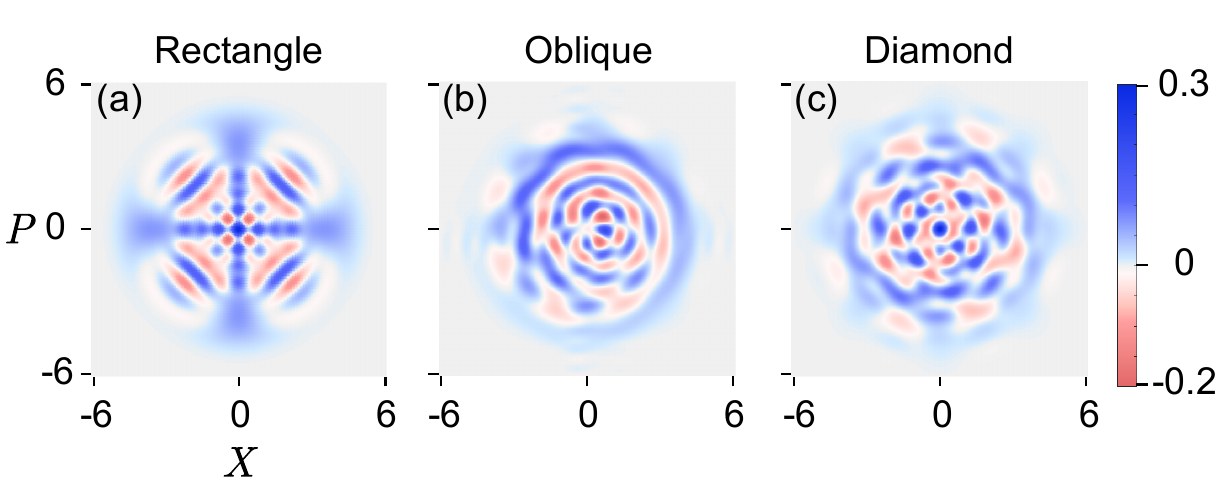}
\caption{The $X$-$P$ Wigner function of the logical state $|+\rangle_L$ of three generalized NP codes with the same mean excitation $\bar{n}_{\mathrm{code}}=6$. (a) The binomial code $[s=4;~f=0;~K=3]$. (b) The O-NP code $[s=1;~f=1/4;~r=0]$. (c) The D-NP code $[s=2;~f=1/2;~r=0]$. Here, the parameters of these logical states are defined in Table~\ref{table1}, and the dual logical state $|-\rangle_L$ can be obtained via an additional phase rotation $\hat{R}(\pi/s)$.}
\label{fig2}
\end{figure}
\smallskip{}

\noindent \textbf{Implementation of generalized NP codes}

\noindent To illustrate generalized NP codes with the familiar quadrature representation, the following discussion will return from the NP representation to the continuous variables representation, and the traditional $X$-$P$ Wigner functions of the logical state $|+\rangle_L$ for generalized NP codes are shown in Fig.~\ref{fig2}. Since $\mathbf{n}_x$ direction of the logical state $|+\rangle_L$ of O-NP and D-NP codes are not orthogonal to the canonical phase, their quadrature representation is non-local in the canonical phase, in contrast to the R-NP codes.

The quantum error correction using rotation-symmetric (R-NP) codes, such as binomial codes and cat codes, has been extensively investigated in both theory and experiment. However, how to realize quantum error correction using O-NP and D-NP codes, as shown in Figs.~\ref{fig2}(b) and (c), respectively, has rarely been studied~\cite{Li2021}. Here, we propose a standard method to remove an NP displacement error $\hat{\mathcal{D}}(\mathbf{n}_e)$ by using the number-phase vortex effect of the O-NP codes. 

As shown in Fig.~\ref{fig1}(e), the number-phase vortex effect of generalized NP codes $(s,f)$ can be described as     
\begin{align}
   \hat{\Sigma}_l|\psi\rangle_L=\exp(-\frac{ifl^2\pi}{2s})\hat{R}(-\frac{lf\pi}{s^2})|\tilde{\psi_l}\rangle_L.
\end{align}
Here, $|\psi\rangle_L=a|0\rangle_L+b|1\rangle_L$ is an arbitrary logical qubit state, and $|\tilde{\psi_l}\rangle_L$ (unnormalized) is the number-shifted qubit state with the Fock-basis replacement $|sn\rangle\rightarrow|sn-l\rangle$. Due to the NP vortex effect, every number-shift error of logical qubit state will lead to a discrete phase rotation with an angle $f\pi/s^2$. In other words, this discrete phase rotation becomes syndromes of number-shift errors when the rotated code words can be correctly distinguished. Note that the syndromes formed by rotated codewords generally have a small non-zero overlap, which is different from the exactly orthogonal number parity. Nonetheless, such an overlap will tend to vanish as long as the phase uncertainty of codewords becomes small enough.  

\begin{figure*}[tbp]	
\centering
\includegraphics[width=0.75\linewidth]{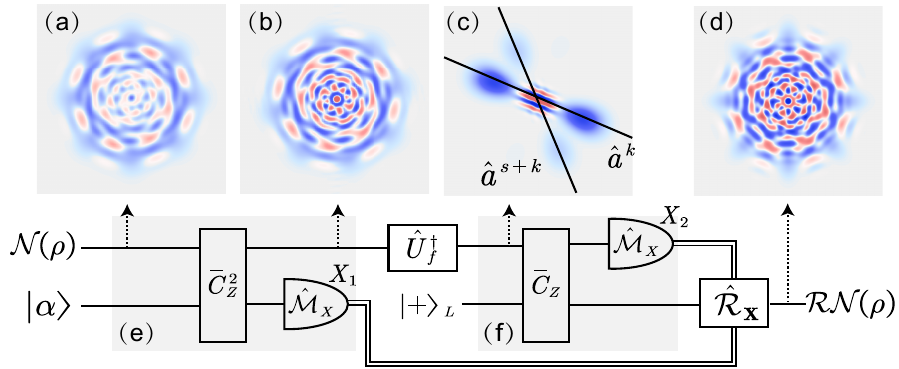}
\caption{Circuit representation of the standard QEC procedure of generalized NP codes. (a)-(d) are the Wigner functions of noisy NP codes in different error correction stages. Here, we choose the logical state $|+\rangle_L$ of a D-NP code $[s=2;~f=1/2;~r=-0.1;~\bar{n}_{\mathrm{code}}=9]$ as an example. The noisy state $\mathcal{N}(\hat{\rho})$ [(a)] has an overlap $\sim0.37$ with the ideal logical state, which is obtained by the evolution in the error model Eq.~\eqref{eq16} with $\gamma t=0.1$ and $\kappa t=0.01$; (b) is obtained with the single-photon loss event of code state is identified by the number parity measurement. The Wigner function (c) is obtained after performing the interface gates $\hat{U}^\dagger_f$, which is back to the R-NP structure, and the shallower peaks are present due to the NP vortex effect. The state (d), corrected via a perfect QEC cycle, has an overlap $\sim0.97$ with the ideal logical state. (e) and (f) are the circuit representations of the modular number-parity measurement and the teleportation-based QEC, defined in Eqs.~\eqref{eq12} and \eqref{eq14}, respectively. The double lines indicate the transmission of classical measurement outcomes for the recovery feedback. } 
\label{fig3}
\end{figure*}

Especially, let us consider the QEC only based on the NP vortex effect of O-NP codes ($s=1$), whose codeword $|\pm\rangle_L$ only has a trivial rotation symmetry $\hat{S}_z=e^{i2\pi\hat{n}}$. Therefore, the identification of number-shift errors through the number-parity measurement is unavailable. Here we introduce the recovery implementation for noisy O-NP codes based on the one-step teleportation~\cite{Knill2005a,Knill2005b,Dawson2006,Grimsmo2020}:
\begin{equation}
\centering
   \includegraphics[width=6.2cm]{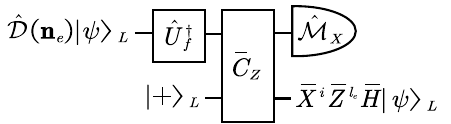}
   \label{eq12}
\end{equation}
where the irrelevant parameters $s=1$ and $\theta$ are omitted, and $\mathbf{n}_e=(l_e,\phi_e)$ is an arbitrary vector in the region Eq.~(\ref{eq7}). $\bar{X},~\bar{Z},~\bar{H}$ are logical Pauli $X$ gate, Pauli $Z$ gate, and Hadamard gate, respectively. The quantum gate 
\begin{equation}
 \bar{C}_Z(s_1,s_2)\coloneqq\exp(-i\frac{\pi}{s_1 s_2}\hat{n}_1\otimes\hat{n}_2)  
 \label{eq13}
\end{equation}
is the controlled-phase gate of two NP codes described by the parameters $(s_1,f_1)$ and $(s_2,f_2)$, respectively.

When a noise-free ancilla mode is initialized at the dual basis $|+\rangle_L$, the $\bar{C}_Z$ gate can entangle the data mode and the ancilla mode via the mechanism $\bar{C}_Z|s_1m\rangle\otimes|s_2n\rangle=(-1)^{mn}|s_1m\rangle\otimes|s_2n\rangle$, where the negative sign is present only when integers $m$ and $n$ are both odd. Due to the NP vortex effect, unknown number shifts will lead to the rotation $\hat{R}(-l_ef\pi)$ of the codespaces, forming as different error subspaces, and the phase difference between two adjacent error subspaces is $\pi/d_N$, thereby the pure phase rotation error with $|\phi_e|<\pi/(2d_N)$ can be treated as a distinguishable disturbance.

Further, a canonical phase measurement $\hat{\mathcal{M}}_X$ is performed on the data mode to identify the rotated dual basis $\hat{R}(-l_ef\pi)|\pm\rangle_L$, which not only gives the error syndromes of the number-shifts $l_e$, but also indicates the projection into $|+\rangle_L$ or $|-\rangle_L$ and simultaneously teleports the quantum state from the data mode to the noise-free ancilla mode with additional logical operations $\bar{X}^i\bar{Z}^{l_e}\bar{H}$. Here $i=0,1$ refers to the dual basis $``+",``-"$, respectively. This teleportation is successful when the rotated dual basis of the data mode can be correctly distinguished from the presence of noise. Note that the noisy qubit state $\hat{\mathcal{D}}(\mathbf{n}_e)|\psi\rangle_L$ requires to be converted into the R-NP structure via the interface $\hat{U}^\dagger_f$ to obtain the minimal canonical phase uncertainty of the dual basis $|\pm\rangle_L$  before the canonical phase measurement.

Due to the code distance of phase variable $d_{\phi}=\pi$, the identifiable number-shift error $l_e$ is up to $q-1$ and phase rotation error $| \phi_e|<\pi/(2q)$. Such a QEC ability is similar to the R-NP codes with the code distance $d_N=s>1$. However, instead of the R-NP codes that use the variation of number parity to identify the number-shift error, the NP vortex effect is an independent quantum resource for correcting the number-shift errors with the syndromes of discrete phase rotation. 

We then discuss the QEC implementation of the generalized NP codes with non-trivial parameters $s$ and $f$, which only requires adding a modular number-parity measurement before the teleportation circuit Eq.~(\ref{eq12}). The modular number-parity measurement is based on the mechanism $\bar{C}_Z^2|sn-k\rangle\otimes|\alpha\rangle=|sn-k\rangle\otimes|\alpha e^{-i2\pi k/s}\rangle$, where $\bar{C}_Z^2$ is the doubled phase-controlled gate and $k=(l_e~\mathrm{mod}~s)$ represents the detectable number-parity.The circuit representation of modular number-parity measurement is:  
\begin{equation}
\includegraphics[width=6.4cm]{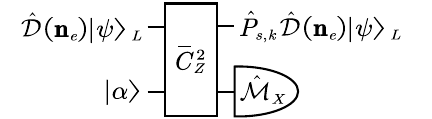}
\label{eq14}
\end{equation}
The $k<s$ number shift of the data mode will cause a phase rotation with the angle $2\pi k/s$ of the coherent state stored in the ancilla mode. This phase rotation can be captured by a canonical phase measurement. The measurement result of the ancilla mode, associated with the measurement operator $|\alpha e^{-i2\pi k/s}\rangle\langle\alpha e^{-i2\pi k/s}|$, corresponds to the application of the projective operator $\hat{P}_{s,k}=\sum_{n=1}^\infty|ns-k\rangle\langle ns-k|$ on the data mode. The modular number-parity measurement can be exact in principle because the phase uncertainty of the coherent state can arbitrarily tend to be vanished as long as the amplitude $\alpha$ is large enough.

The standard QEC implementation circuit of generalized NP codes is illustrated in Figs.~\ref{fig3} (e) and (f). Based on the outcomes $\mathbf{X}=(X_1,X_2)$ of the modular number-parity measurement and the phase measurement in teleportation, we can recover the noisy qubit states $\mathcal{N}(\rho)$ after a recovery operation
\begin{equation}
\hat{\mathcal{R}}_{\mathbf{X}}=\bar{H}\bar{Z}^m\bar{X}^{i}\bar{Z}^{\frac{k}{s}}.
\label{eq15}
\end{equation}
Here the integer $k\in[0,s-1]$ is given by the outcome $X_1=-2\pi k/s$, while the integer $i$ and $m$ are obtained from the outcome $X_2-\phi_{k,s}=(\pi/s)^i-mf\pi/s+\phi_e$. The global phase $\phi_{k,s}=-fk\pi/s^2$ is fixed by the previous number-parity measurement, and $|\phi_e|<\pi/(2d_N)$ is the correctable phase rotation noise. Due to the code distance $d_{\phi}=\pi/s$ of the generalized NP codes, the detectable integer $m\in[\lfloor L/s\rfloor-q+1,\lfloor L/s\rfloor]$ and $k$ determine the amplitude of the number shift $l_e=ms+k$. Note that the recovery operation involves fractional order of logical $\bar{Z}$ gate, whose implementation can be easily achieved via a simple phase rotation operation $\bar{Z}^{\frac{k}{s}}=\hat{R}(i\pi k/s^2)$.


\smallskip{}

\noindent \textbf{QEC performance of generalized NP codes}

\noindent Based on the above implementation, we investigate the QEC performance of the generalized NP codes by considering a noise channel $\mathcal{N}(\hat{\rho})$ that is the solution of the master equation
\begin{equation}
\frac{\partial\hat{\rho}(t)}{\partial t}=\gamma\mathcal{L}[\hat{a}]\hat{\rho}(t)+\kappa\mathcal{L}[\hat{n}]\hat{\rho}(t),
\label{eq16}
\end{equation}
with an integral time $t$, and $\mathcal{L}[\hat{A}]\hat{\rho(t)}=\hat{A}\hat{\rho(t)}\hat{A}^\dagger-1/2\rho(t)\hat{A}^\dagger\hat{A}-1/2\hat{A}^\dagger\hat{A}\rho(t)$. Such an error model can describe the energy decay and quantum dephasing of a bosonic field stored in a cavity or transmitting in a waveguide. For the integral time $t$ small, the physical single-photon loss and dephasing operator can be described in terms of NP displacement operators as
\begin{align}
\label{eq17}
\sqrt{\gamma t}\hat{a}e^{-\frac{\gamma t}{2}\hat{n}}&\approx\int_{-\pi}^{\pi}d\phi~\frac{\sqrt{\gamma t}\exp(\frac{i\phi}{2})}{4\sqrt{\pi}(\gamma t/2+i\phi)^{3/2}}\hat{\mathcal{D}}(1,\phi),\\
\sqrt{\kappa t}\hat{n}&\approx\frac{1}{2i}[\hat{\mathcal{D}}(0,\sqrt{\kappa t})-\hat{\mathcal{D}}(0,-\sqrt{\kappa t})].
\end{align}
The left side of Eq.~\eqref{eq17} is actually the first-order Kraus operator of pure-loss channel (i.e., Eq.~\eqref{eq16} with $\kappa=0$)~\cite{li2017,Albert2018}. Note that the modulus of the superposition coefficient on the right side of Eq.~\eqref{eq17} is related to a Lorentzian function with a linewidth $\gamma t$. If the square root term $\sqrt{\gamma t}$ is retained while the higher-order term $\gamma t$ is ignored, an unphysical representation of single-photon loss can be obtained, where the expansion coefficient diverges when $\phi=0$. The error rates must satisfy $\gamma t,\kappa t\ll 1$ to guarantee sufficiently small NP displacements of the codewords, thereby ensuring effective QEC. Additionally, the analogous representation for the single-photon gain operator can be straightforwardly obtained as the complex conjugation of the single-photon loss.

The noisy code states then need to be sent to the QEC circuit to remove the noise and recover the logical information. As an example, we illustrate the QEC of a noisy D-NP code to demonstrate how the error correction works, as shown in Fig.~\ref{fig3}. After the recovery process, the QEC performance of the generalized NP codes can be qualified by the channel fidelity~\cite{Peimpell2005}
\begin{equation}
F=\frac{1}{4}\sum_{\mathbf{X},j}|\mathrm{Tr}(\hat{\mathcal{R}}_{\mathbf{X}}\hat{N}_j)|^2,
\end{equation}
where $\hat{N}_j$ is the Kraus operator of the noise channel $\mathcal{N}$, and the Kraus-operator decomposition of the noise channel $\mathcal{N}$ is given in the supplementary Note 4. 

In principle, the generalized NP codes with the same code distance $d_N=sq$ should perform similarly when the phase uncertainty is small enough. It indicates the mean excitation of codes $\bar{n}_{\mathrm{code}}$ is large, and thus the bit-flip error is dominant. However, the performance of NP codes will become different when the phase-flip error is non-negligible with the small $\bar{n}_{\mathrm{code}}$. This performance difference comes from both the different NP lattice structures and the Fock-amplitude series $\{\theta_n\}$. 

\begin{figure}[tbp]	
\centering
\includegraphics[width=0.95\linewidth]{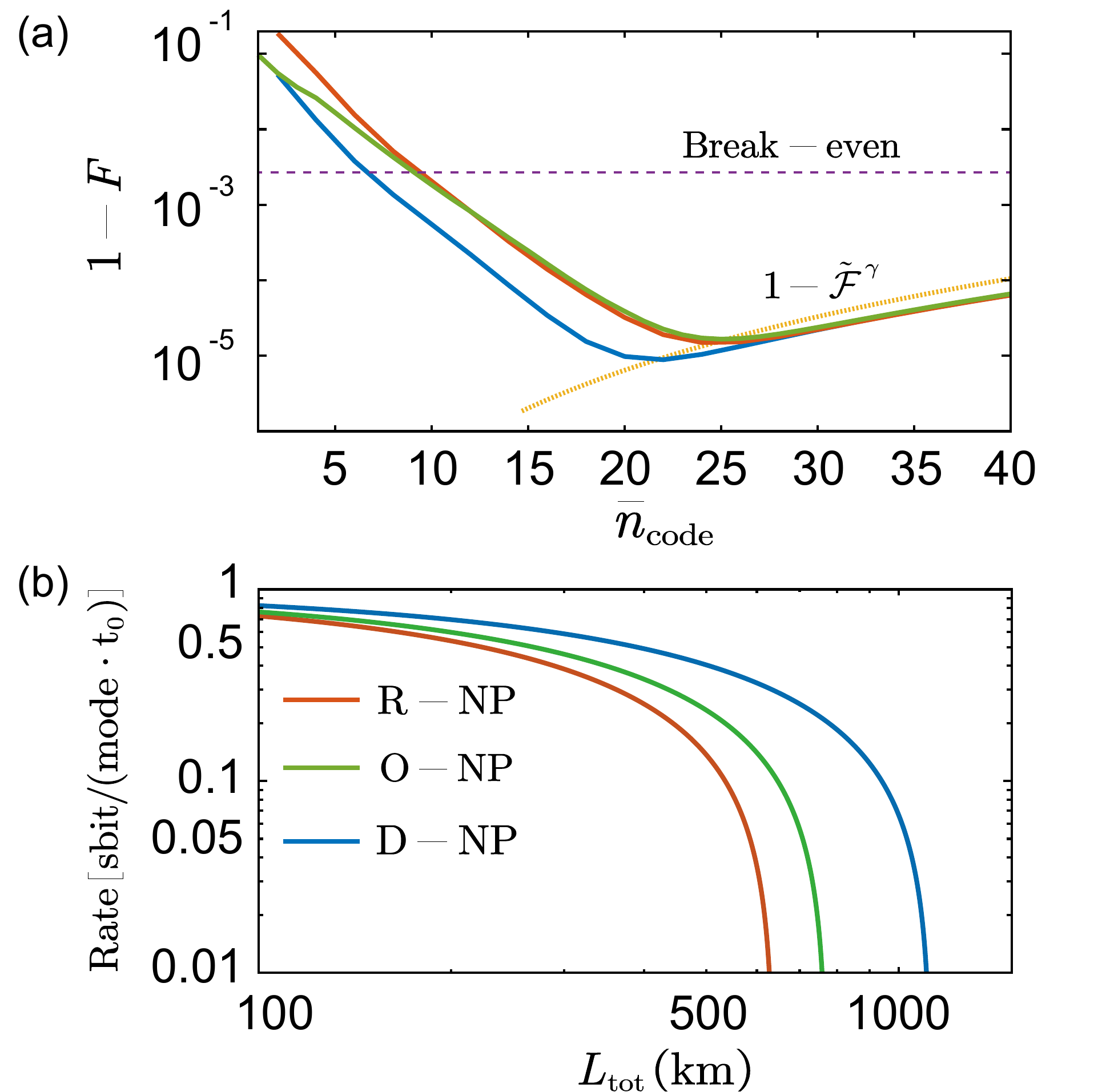}
\caption{QEC performance comparison of the different generalized NP codes. (a) Optimized channel infidelity as the function of the mean excitation $\bar{n}_{\mathrm{code}}$ of generalized NP codes which have the same code distance $d_N=4$. Here, we choose the error rate $\gamma t=0.5\%$ and $\kappa t=0.1\%$. The dotted orange line corresponds to the data obtained from Eq.~\eqref{eq20}, and the break-even line is estimated using Fock states $|0\rangle$ and $|1\rangle$ for coding. (b) Optimized SKRPM of generalized NP codes with the fiber coupling loss rate $\epsilon=1\%$ and the dephasing error rate $\Gamma_{\phi}=0.1\Gamma$. Here, the y-axis label is the generation rate of secure bits per mode. $t_0$ is the gate operation time taken for QEC, and $1/t_0$ is the raw key generation rate~\cite{sreraman2014}.}
\label{fig4}
\end{figure}

To compare the performance of NP codes with different NP lattice structures, we numerically evaluate the QEC channel infidelity as a function of the mean code excitation $\bar{n}_{\mathrm{code}}$, shown in Fig.~\ref{fig4} (a). As an example, here we focus on the binomial code, the O-NP code, and the D-NP code, whose Fock amplitudes are both selected as $|\theta_n\rangle=\langle n|\alpha,r\rangle$. Among cat and binomial codes, we selected binomial codes as the representative R-NP codes because previous research data in Refs.~\cite{Albert2018,Grimsmo2020} indicate that binomial codes exhibit better QEC performance than cat codes under the same conditions. These NP codes have the same code distance $d_N=4$, and the squeezed parameter $r$ of the O-NP and D-NP codes is optimized to minimize the channel infidelity. The numerical results show that the binomial codes and the O-NP codes have a similar performance, while the D-NP code has an outstanding performance when the mean code excitation $\bar{n}_{\mathrm{code}}$ is small. This phenomenon indicates that correcting number-shift errors using only the number degree (R-NP codes) or the phase degree (O-NP codes) has similar QEC performance; however, correcting number-shift errors using a hybrid number-phase degree can enhance QEC performance.   

As expected, the three types of NP codes have similar performance when $\bar{n}_{\mathrm{code}}$ is large enough, as shown in Fig.~\ref{fig4}(a). In such a region, the phase-flip error is negligible, and the dominant bit-flip rate can be estimated as
\begin{align}
    1-\tilde{\mathcal{F}}^{\gamma}\approx \sum_{l=1}\frac{2(\bar{n}_{\mathrm{code}}\Gamma)^{ld_N}e^{-\bar{n}_{\mathrm{code}}\Gamma}}{(ld_N)!}.
    \label{eq20}
\end{align}
This can be derived from the QEC matrix $M$ which is defined as $M_{[j\mu],[k\nu]}=\langle\mu|\hat{N}^\dagger_{j}\hat{N}_k|\nu\rangle$. It is based on the near-optimal performance~\cite{Zheng2024} of NP codes, and the detailed process can be found in the supplementary Note 5. For a given excitation-loss rate $\Gamma=1-e^{-\gamma t}$, the dominant bit-flip error rate can be suppressed by increasing the code distance $d_N$.

On the other hand, the ability of NP codes against pure-dephasing entirely depends on the Fock amplitude $\theta_n$. Though the analytical infidelity expression caused by pure dephasing for different NP codes is not trivial, they have similar performance when the phase uncertainty is very small. We can find an approximate lower bound of the QEC performance of NP codes against the pure-dephasing noise, which is obtained from the ideal NP codes, as
\begin{align}
    (1-\mathcal{F^{\kappa}})_{\min}\approx\frac{(8\pi s^2\kappa t)^{\frac{1}{2}}}{\pi^2}\exp(-\frac{\pi^2}{8s^2\kappa t}).
    \label{eq21}
\end{align}
Here, the lower bound is dominant by a factor $e^{-\frac{\pi^2}{8s^2\kappa t}}$, and the remainder is an effective approximation when the dephasing rate $\kappa t\ll1$. Though the lower bound unfortunately does not exist a simply exact expression, it already shows the key physical mechanism that the pure dephasing rate $\kappa t$ will be enhanced by the parameter $s$ with a quadratic scaling. This indicates that the presence of pure dephasing prevents NP codes from working with an excessively large $s$. Under such a condition, phase-flip errors become the dominant error mechanism, thereby compromising the QEC performance of NP codes.

\smallskip{}
\noindent \textbf{One-way quantum communication}
\noindent The generalized NP codes, especially the well-known R-NP codes, have been experimentally demonstrated to have many applications. For example, the cat codes with QEC can extend the lifetime of a qubit~\cite{Ofek2016}, and the binomial codes with QEC can protect the entanglement between two qubits~\cite{Cai2024a}. 

The generalized NP codes with repetitive QEC are also promising to execute long-distance quantum communication based on the quantum repeaters (QRs)~\cite{Briegel1998,Jiang2009,Lo2014,Takeoka2014,li2024,Rozpedek2021,Wu2022}. One round QEC should correct the transmission loss $1-e^{-\tilde{L}_0}$ and the the fiber coupling loss $\epsilon$, where $\tilde{L}_0=L/L_{att}$ is the dimensionless repeater spacing, and $L_{att}=20~\mathrm{km}$ for optical fiber~\cite{Kunz2018}. In addition, the energy attenuation in nonlinear optical fibers with Kerr medium can lead to pure phase rotation errors of the quantum states, which are usually ignored by previous studies~\cite {li2017,Kunz2018}. In the QEC process shown in Fig.~\ref{fig3}, the imperfect quantum gates and measurements also result in dephasing errors~\cite{Lescanne2020,Joshi2021}. Given that the commutation of all dephasing errors with both the interface gate and the controlled-phase gate, these errors can be theoretically treated as occurring before the ideal QEC process. Therefore, the excitation-loss and dephasing errors can be collectively modeled by the Eq.~(\ref{eq16}), and the excitation-loss rate of one instance QEC is $\Gamma=1-e^{-\tilde{L}_0}+\epsilon$, while the secondary dephasing error rate is modeled as $\Gamma_{\phi}=h\Gamma$ with $h<1$. 

For the communication distance as far as possible, the NP codes should be optimized to minimize the error accumulation rate
\begin{align}
    \tau(s,f,\theta,\Gamma,\Gamma_\phi)=\frac{1-F}{\tilde{L}_0}.
\end{align}
We quantify the QR performance of three types of NP codes by considering the quantum key distribution. In Fig.~\ref{fig4} (b), we compare the optimized secure key rate per mode (SKRPM) of three NP codes, where the fiber coupling loss $\epsilon=1\%$, and the dephasing error rate is selected as $\Gamma_{\phi}=0.1\Gamma$. The three types of NP codes yield an SKRPM $>0.01$ for a near-thousand-kilometer quantum communication. As expected, the binomial codes and the O-NP codes have similar performance for quantum communication, while the D-NP codes have an outstanding performance. The slight advantage of O-NP codes over binomial codes arises from their superior tolerance to pure rotation errors when the code distance $d_\phi=\pi/s$ of the binomial code is small enough. Moreover, we also simulate their QRs performance with $\Gamma_{\phi}=0.05\Gamma$, where the communication distance is significantly extended as the dephasing error rate $\Gamma_{\phi}$ is further suppressed, and the behavior of performance remains unchanged. The detailed numerical results are summarized in supplementary Table 1, which contains the Refs.~\cite{Wang2005,Scarani2009}.  

\smallskip{}

\noindent \textbf{Experimental feasibility and limitations}

\noindent The simulated QEC performance of generalized NP codes presented above assumes an ideal, error-free cycle, demonstrating their theoretical potential for practical applications. For experimental implementation, superconducting quantum circuits currently represent the most viable platform due to their advanced control capabilities and scalability. Here, we provide a comprehensive analysis of the experimental requirements and practical limitations for realizing these codes in state-of-the-art superconducting quantum circuit systems.

In experiments, the encoded bosonic mode is usually stored in a high-quality microwave cavity with a lifetime $T_1 \sim 1~ms$~\cite{milul2023superconducting}, and a coupled two-level ancilla transmon with $T_1\sim0.1~ms$ is introduced to manipulate the bosonic mode~\cite{Ni2023}. In our case, the teleportation-based QEC protocol for generalized NP codes requires the operations set 
\begin{equation}
  \{\mathcal{P}_{s}(|+\rangle),~\bar{H},~\bar{Z}(\beta),~\hat{U}_f,~\bar{C}_Z,~\hat{\mathcal{M}}_X\}  
  \label{eq23}
\end{equation}
on bosonic modes. Here $\bar{Z}(\beta)=\mathrm{diag}(1,e^{i\beta})$ represents the arbitrary rotation of generalized NP codes $(s,f)$ around the Z-axis in the logical level, which includes the phase gates $\bar{S}=\mathrm{diag}(1,i)$ and $\bar{T}=\mathrm{diag}(1,e^{i\pi/4})$. The combination of $\bar{Z}(\beta)$ and Hadamard gate $\bar{H}$ can realize universal rotation on a single logical qubit. In addition, $\mathcal{P}_{s}(|+\rangle)$ denote the preparation for the logical state $|+\rangle_L(s,0)$ of R-NP codes, then the corresponding logical state $|+\rangle_L(s,f)$ of generalized NP codes can be obtained via the interface gate $|+\rangle_L(s,f)=\hat{U}_f|+\rangle_L(s,0)$.

Among these required operations, the crucial challenge is the experimental implementation of destructive canonical phase measurement $\hat{\mathcal{M}}_X$, which is used to identify error syndromes and teleport the quantum state. A related experiment has been demonstrated for a qubit encoded by Fock states $|0\rangle$ and $|1\rangle$ in superconducting quantum circuits~\cite{Martin2020}. However, the discrimination for rotated codewords of NP codes still needs to extend this experiment beyond the single-photon regime. In the absence of implementing a canonical phase measurement, heterodyne detection is a convenient method to readout the phase of an unknown signal with lower precision. The heterodyne is the well-know Gaussian POVMs with element $\hat{E}(\alpha)=\pi^{-1}|\alpha\rangle\langle\alpha|$, which projects the bosonic mode onto a coherent state $|\alpha\rangle$ and the phase is estimated as $\mathrm{Arg}(\alpha)$, while the number information $|\alpha|$ is wasted. In addition, a more practical limitation is the finite readout efficiency of microwave cavities~\cite{pfaff2017controlled,walter2017rapid,Martin2020}. A detailed discussion of the reduced QEC performance in R-NP codes resulting from imperfect phase measurements is presented in Ref.~\cite{hillmann2022performance}. 

The above phase measurements are independent of the encoding, offering generality for NP codes with different $\{\theta_n\}$. When the mean code excitation $\bar{n}_{\mathrm{code}}$ is large enough, the intrinsic canonical phase uncertainty of the codeword $|\pm\rangle_L(s,0)$ will tend to vanish, thereby the rotated codewords can be effectively discriminated via the canonical phase measurement or heterodyne detection. However, in the low-excitation regime, the intrinsic canonical phase uncertainty of the codewords will limit the phase measurement accuracy, even if the phase measurement is perfect. These mechanisms are intuitively reflected in the variation of QEC performance and KL cost function for NP codes with $\bar{n}_{\mathrm{code}}$, as shown in Fig.~\ref{fig4}(a) and in supplementary Figure 2, respectively.

On the other hand, a technology proposed by our previous work~\cite{Cai2024b} can serve as an alternative phase measurement for specific NP codes with $\theta_n=\langle n| \alpha,r\rangle$. The noisy NP codes with such a Fock-amplitude $\theta_n$ behave as a mixture of rotated coherent-like states due to the NP vortex effect, following the interface gate operation, as shown in Fig.~\ref{fig3}(c). Coincidentally, this technology has achieved the unambiguous discrimination of up to six coherent states ($|\alpha|\leq2$) uniformly distributed on a phase circle, corresponding to the required phase measurement for NP codes with $d_N=3$. It is a potential way to unambiguously identify error subspaces formed by rotated coherent-like states with a low excitation, and the measurement time is around $\log_2(2d_N)~\mu \rm{s}$.   

Additionally, a key step in implementing generalized NP codes is the state preparation of logical states $|+\rangle_L (s,0)$ for R-NP codes. It is not a trivial task to prepare such a state with a large Fock space interval $s$. In circuit QED, a universal method for preparing an arbitrary state of a target cavity mode is available in principle. It involves the strong dispersive interaction between the cavity mode and an ancilla transmon, combined with quadrature displacement operations performed on the cavity mode~\cite{HeeresPRL2015,Krastanov2015,Heeres2017,MA2021Bosonic}. In experiments, this method performs well for preparing high-fidelity logical states with several photons in $\sim1~\mu \rm{s}$~\cite{Ofek2016,Ni2023}. However, preparing logical states with good phase distinguishability for high-order R-NP codes remains challenging via this method, as such states typically contain tens of photons. This arises because the process involves long operation times and complex controls, where the photon decay of the short-lifetime ancilla transmon introduces more noise, leading to errors in the state preparation.

The transformation from the R-NP codes $(s,0)$ to generalized NP codes $(s,f)$ is based on the interface gate, as shown in Eq.~\eqref{eq9}, which is also frequently used in teleportation-based QEC. Its implementation requires controlled self-Kerr interaction $H=K\hat{n}^/2$ for the data mode. This technology has been achieved in a Kerr-tunable superconducting resonator~\cite{he2023fast}, where the self-Kerr interaction strength $K/2\pi$ can be adjusted from $-5$ MHz to $6$ MHz, and the special logical state $|+\rangle_L(1,f)=\hat{U}_f|\alpha\rangle$ of the O-NP code (with $f=1/2,~1/3,~1/4$ and $\bar{n}_{\mathrm{code}}=2$) is prepared within a gate time of $\sim 0.1 \mu \rm{s}$. However, this experimental work was conducted in the strong dissipation regime ($\gamma/K_{\max}\approx0.1$). Therefore, the realization of interface gates still requires the extension of the data mode lifetime to ensure that noise caused by photon loss during operations is negligible ($\gamma/K_{\max}\ll1$). In a recent experiment~\cite{hajr2024high}, a comparable superconducting resonator device achieved a lifetime of $T_1\sim40~\mu \rm{s}$, presenting a promising platform to implement the feasible interface gate with an estimated $\gamma/K_{\max}$ ratio of $\sim2.5\times10^{-3}$.

In Fig.~\ref{fig3}, we present a QEC scheme only using bosonic modes, which is expected to perform QEC under bosonic-level noise and is promising for realizing fault-tolerant quantum computation based on bosonic modes. However, such a scheme requires entangling gates implemented by cross-Kerr interaction between two bosonic modes, which is typically weak ($K_c/2\pi\sim10$ kHz) for two cavities in the current experiments. Therefore, Fig.~\ref{fig3} represents a long-term goal to realize QEC with all bosonic elements. For the current experimental considerations, we introduce a general method to conditionally implement universal logical rotation of an NP code, and the logical entangling $\bar{C}_Z$ gate of any two NP codes stored in different cavities. It is based on a gadget that a generalized NP code $(s,f)$ stored in a data mode entangles an ancilla transmon prepared in $|+\rangle
_q=(|e\rangle+|g\rangle)/\sqrt{2}$ through the gate
\begin{equation}
\bar{C}_q(s)\coloneqq\exp(-i\frac{\pi}{s}\hat{n}\otimes|e\rangle\langle e|),    
\end{equation}
where $|e\rangle$ and $|g\rangle$ denote the excited state and the ground state of the ancilla transmon, respectively. This gate serves the same role as the controlled phase gate $\bar{C}_Z(s,1)$ between two NP codes $(s,f)$ and $(1,0)$ described by Eq.~\eqref{eq13}, and here the transmon replaces the trivial NP code $(1,0)$ with codewords are Fock state $|0\rangle$ and $|1\rangle$. In contrast to the direct cross-Kerr interaction between two cavities, the dispersive interaction $\chi\hat{n}_1\otimes|e\rangle\langle e|$ between a cavity and a transmon is more natural and strong ($\chi/2\pi\sim4$ MHz)~\cite{Ni2023}.

By using the gadget that the NP code $(s,f)$ entangles the ancilla transmon through the gate $\bar{C}_q$, one can conditionally realize the logical rotation $\bar{Z}(\beta)$ through the teleportation circuit
\begin{equation}
\centering
 \includegraphics[width=8cm]{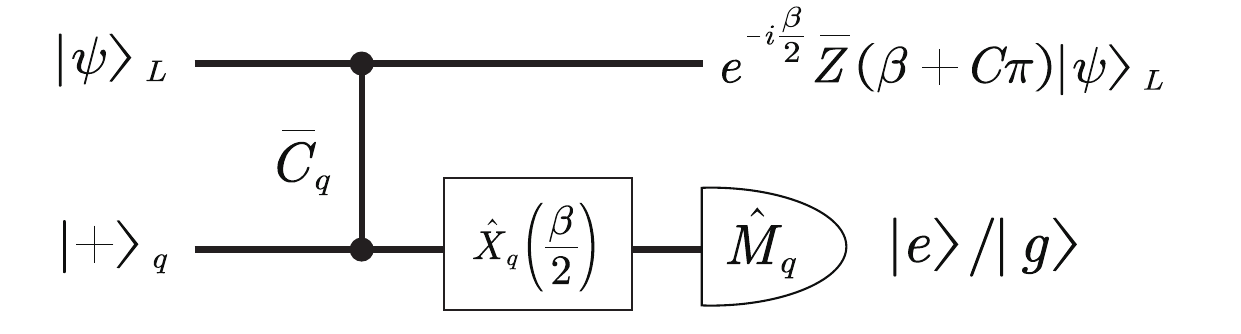} 
 \label{eq26}
\end{equation}
Here $\hat{X}_q(\beta/2)\coloneqq\exp(-i\beta\hat{\sigma}_x/2)$ is the single-qubit rotation of the transmon and $\hat{\sigma_x}=|g\rangle\langle e|+|e\rangle\langle g|$. After measuring $|e\rangle$ and $|g\rangle$ basis of the transmon via $\hat{M}_q$, the logical rotation $\bar{Z}(\beta)$ can be conditionally applied for the dada mode. Here an additional logical rotation $\bar{Z}(C\pi)=\bar{Z}^C$ with $C=0$ or $C=1$ arises depending on the measurement outcome: $|g\rangle$ or $|e\rangle$, respectively. The gate operation time of $\bar{Z}(\beta)$ is limited by the implementation of $\bar{C}_q$, which can be estimated as $1/\chi\sim0.1~\mu \rm{s}$.

To achieve the universal logical rotation of the generalized NP codes $(s,f)$, it is necessary to implement the logical Hadamard gate. This task can be conditionally realized through the cascaded teleportation circuit
\begin{equation}
\centering
 \includegraphics[width=7.8cm]{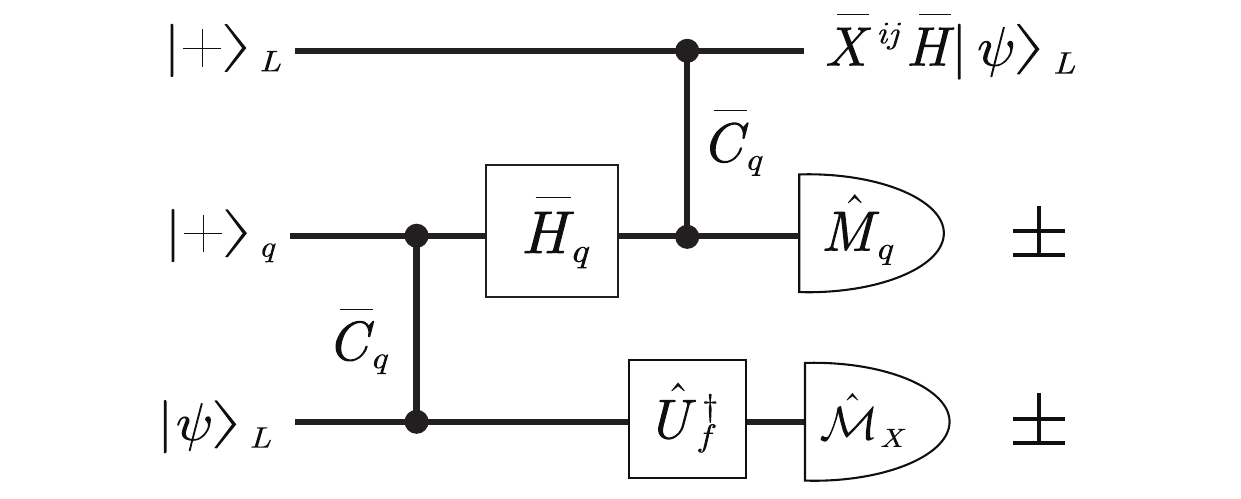} 
 \label{eq27}
\end{equation}
Here $\bar{H}_q$ is the Hadamard gate of the transmon qubit. This circuit can be interpreted as follows: the measurement for basis $|\pm\rangle_q$ conducted on the middle rail effectively induces a controlled-phase gate $\bar{C}_Z$ between the top and bottom rails. Although this method induces a conditional logical Pauli operation $\bar{Z}^i$ (not shown) on the bottom rail, where $i=0,~1$ corresponds to the measurement outcome "$+$" and "$-$", respectively, it represents a more practical approach to realizing the entangling $\bar{C}_Z$ gate than directly manipulating the cross-Kerr interaction between two cavities. This controlled-phase gate protocol is also limited by two successive gates of $\bar{C}_q$, resulting in an estimated gate time $\sim 0.2~\mu \rm{s}$.   

Then, the logical Hadamard gate $\bar{H}$ can be conditionally implemented via identifying the basis $|\pm\rangle_L$ of the bottom rail via the phase measurement. This scheme is equivalent to Eq.~\eqref{eq12} in the absence of noise. A logical operation $\bar{X}^j$ is present after the Hadamard gate, where$j=0,~1$ corresponds to the measurement outcome "$+$" and "$-$", respectively. The operation time of this gate protocol is constrained by the phase measurement, which takes approximately $2\sim 4~\mu \rm{s}$ for corresponding code distance $d_N=2\sim10$ using unambiguous state discrimination technology~\cite{Cai2024b} or even less for heterodyne detection.    

\begin{table}[thb]
\centering
\caption{Reference durations $T$ for required QEC operations in superconducting circuit experiments. These operations are, in order from left to right, state preparation $\mathcal{P}_{s}(|+\rangle)$, logical Hadamard gate $\bar{H}$, arbitrary logical $\bar{Z}$ rotation $\bar{Z}(\beta)$, interface gate $\hat{U}_f$, logical controlled-phase gate $\bar{C}_Z$, and phase measurement $\hat{\mathcal{M}}_X$.}
\renewcommand{\arraystretch}{2}
\begin{tabular}{m{1.3cm}<{\centering} m{1.3cm}<{\centering} m{1.3cm}<{\centering} m{0.8cm}<{\centering} m{0.8cm}<{\centering} m{0.9cm}<{\centering} m{1.3cm}<{\centering}}
\hline\hline
Operation & $\mathcal{P}_{s}(|+\rangle)$ & $\bar{H}$ & $\bar{Z}(\beta)$ &$\hat{U}_f$& $\bar{C}_Z$& $\hat{\mathcal{M}}_X$\\\hline\hline
$T(\sim\mu \rm{s})$  & $1$ & $2\sim4$ & $0.1$ & 0.1&0.2& $2\sim4$ \\\hline
Ref.& \cite{Ofek2016}& \eqref{eq27}&\eqref{eq26} &\cite{he2023fast}&\eqref{eq27}&\cite{Cai2024b}
\\\hline\hline
\end{tabular}
\label{table2}
\end{table}

We summarize the experimental reference durations $T$ for these required operations in Table \ref{table2}. Compared to the lifetimes of the cavity modes and ancilla transmon, these operations are sufficiently fast to render the implementation of generalized NP codes with current technology entirely feasible. Notably, we construct the universal logical rotation for a generalized NP code and the logical controlled-phase gate between two generalized NP codes, both based on quantum teleportation. Such a scheme inevitably introduces conditionally additional Pauli operations. A straightforward solution is to apply corresponding recovery feedback to remove these additional logical Pauli operations based on measurement outcomes, where the logical Pauli Z of NP codes $(s,f)$ can be easily implemented via the phase rotation $\hat{R}(\pi/s)$, while the logical Pauli X operation $\hat{\mathcal{D}}(\mathbf{n}_x)$ of NP codes $(s,f)$ involves multi-photon transitions between Fock states, rendering it difficult to directly realize on bosonic modes.

A more practical approach was proposed in Ref.~\cite{Grimsmo2020}, where fault-tolerant quantum computation using R-NP codes is comprehensively considered. It suggests that rather than explicitly correcting the additional Pauli operations via complex procedures during teleportation-based QEC and quantum computation with magic state injection $\mathcal{P}_{|T\rangle}$ and $\mathcal{P}_{|S\rangle}$, these operations can be marked by each measurement outcome and tracked in software.  

Lastly, due to the photon loss of the ancilla transmon, the required operations for bosonic modes may be faulty. Therefore, we simulate the noisy QEC cycle shown in Fig.~\ref{fig3} with experimental limitations. It demonstrates that the transmon dissipation during the gate operations introduces uncorrectable logical errors, thereby limiting the QEC performance of generalized NP codes. However, the main physical mechanism (number-phase vortex effect) of these codes is not significantly affected by imperfect gate operations. The detailed simulation results can be found in the supplementary Note 7. In a short-term consideration, the photon loss of an ancilla transmon can already be suppressed to the second order by replacing the employed excited state $|e\rangle$ with a higher excited state $|f\rangle$ of a three-level transmon~\cite{rosenblum2018fault}. Furthermore, the fundamental solution for these operational imperfections would require the development of fault-tolerant operations for these codes, which is a challenging but crucial long-term objective. Nonetheless, there are several theoretical frameworks that can be extended to achieve fault-tolerance for generalized NP codes~\cite{Preskill2008,Grimsmo2020,xu2024fault}. 

The QEC of a generalized NP code is based on the one-step teleportation, as shown in Eq.~\eqref{eq12}. Notably, the teleportation between two physical qubits can be extended to achieve fault-tolerance for biased noise by encoding the physical qubits into repetition codes, as proposed in Ref.~\cite{Preskill2008}. For generalized NP codes, after partial number-shift errors are strictly identified by the non-destructive number-parity measurement, the remaining errors are all phase rotation errors that are similar to biased noise. Therefore, a natural ideal is to extend the teleportation between two individual generalized NP codes to fault-tolerant teleportation between two repetition codes encoded by condensed generalized NP codes. For R-NP codes, this method for achieving fault-tolerance has been comprehensively discussed in Ref.~\cite{Grimsmo2020}.   

Another promising scheme for implementing fault-tolerant operations on single-mode bosonic codes with discrete-variable ancillae is based on the path-independent quantum control ~\cite{xu2024fault}. Compared with the previous scheme dedicated to NP codes, this scheme has stronger universality and lower resource consumption, maintaining the hardware efficiency of single-mode bosonic quantum codes.


\smallskip{}

\noindent \textbf{\large{}Discussion}{\large\par}

\noindent So far, we have introduced the definition, error correction, and applications of the generalized NP encoding of a bosonic mode. Here, we discuss and compare their respective advantages in experiments. 

Among the three types of generalized NP codes, the state preparation of O-NP codes is the simplest when the feasible interface gate $\hat{U}_f$ is available. The preparation of code states ($|\pm\rangle_L(1,f)=\hat{U}_f|\pm\alpha,r\rangle$) and modulation of the code distance ($d_N$) for O-NP codes solely necessitate the application of a unitary operator $\hat{U}_f$ on displaced squeezed states. While the state preparation of generalized NP codes $(s,f)$ with a large Fork states interval $s$ is usually difficult, which involves multi-photon transition. Another feature of the O-NP codes is the convenient implementation of approximate logical Pauli $X$ operation $\bar{X}\approx\hat{U}_f\hat{D}[i\pi/(2\alpha)]\hat{U}_f^\dagger$ with the gate infidelity $\propto \exp(-2\alpha^2)$. This is based on the commutation relations between quadrature displacement operators, and the similar gate protocol for two-leg cat codes is proposed in Ref.~\cite{Cochrane1999}. In addition, the logical Pauli $Z$ operation can be easily realized by a phase rotation $\hat{R}(\pi/s)$ for all generalized NP codes $(s,f)$.

Based on above features of O-NP codes, they can be easily implemented in optical systems. The interface gate may be realized by propagating a light field in a Kerr nonlinearity medium, and the generation of optical cat states have been widely studied in experiments~\cite{Ourjoumtsev2007,Sychev2017,Le2018,Hacker2019,Han2023}. Moreover, the most challenging aspect of realizing the one-step-teleportation QEC scheme [Eq.~\eqref{eq12}] for O-NP codes is the correct identification of rotated codewords following the interface gate operation, analogous to the scenario depicted in Fig.~\ref{fig3}(c). In essence, this task constitutes a phase estimation problem for rotated coherent states, and the required phase measurement is also fast-growing in experiments~\cite{Higgins2007,Martin2020,Rodriguez-Garcia2022,Cai2024b}. Consequently, large-sized O-NP encoding of an optical mode is very suitable for executing long-distance quantum communication and interconnection.

On the other hand, the R-NP codes with non-trivial rotation symmetry , such as the cat codes and binomial codes, may suitably implement fault-tolerant quantum computation by encoding the microwave field stored in a superconducting cavity. The code state preparation and universal control for these codes can be achieved by manipulating the microwave cavity with a coupled discrete-variable ancillae. This allows for the strict identification of error syndromes in these codes, followed by the restoration of quantum information. The relevant experimental demonstration and theoretical proposal had a flourishing development in past decades. Moreover, the QEC performance of R-NP codes may be further improved by performing an additional unitary $\hat{U}_{f=1/2}$ on the code states, namely, encoding to be the D-NP codes. The truth that the D-NP codes have higher performance than R-NP and O-NP codes is analogous to which the hexagonal GKP codes have higher performance than square GKP codes. Based on the relative advantages of generalized NP codes in experiments, the faithful quantum computer and the long-distance quantum communication in a quantum network may be achieved using hybrid generalized NP encoding of bosonic modes.

In summary, we construct a novel framework of bosonic encoding, which defines the logical states via discrete number-phase translation symmetry. It is designed to against the number-phase shift error of a bosonic mode, which involves the energy decay and the quantum dephasing. The framework contains many well-known bosonic codes, such as cat codes and binomial codes, and it also guides us to find the O-NP codes and D-NP codes, which can be imagined as number-phase vortex in NP space. Crucially, such a number-phase vortex effect is an independent quantum resource, which paves a new way to correct the number-shift errors using phase degrees. Moreover, the O-NP codes are easily generated and implemented QEC for optical modes, which may allow it to suitably implement long-distance quantum communication in a hybrid quantum network or any application scenarios required to resist the NP-shifts noise in a bosonic mode. We believe that the framework can inspire the experiments and guide more discoveries of error-correcting codes.  

\smallskip{}

\noindent \textbf{\large{}Data availability}{\large\par}

\noindent All data generated or analysed during this study are available within the paper and its Supplementary Information. Further source data will be made available on reasonable request.

\smallskip{}

\noindent \textbf{\large{}Code availability}{\large\par}

\noindent The code used to solve the equations presented in the Supplementary Information will be made available on reasonable request.

\smallskip{}

\noindent \textbf{\large{}Reference}{\large\par}

\clearpage{}

\smallskip{}

\noindent \textbf{\large{}Acknowledgment}{\large\par}

\noindent Z.-L.X. is supported by the National Natural Science Foundation of China (Grant No. 12375025). C.-L.Z. is supported by the Innovation Program for Quantum Science and Technology (Grant No.~2021ZD0300203).

\smallskip{}

\noindent \textbf{\large{}Author contributions}{\large\par}

\noindent The research topic was developed by D.-L.H. and Z.-L.X., and C.-L. Z. and W.C. provided guiding suggestions and improved the theoretical framework. D.-L.H. simulated the data and wrote the draft of the manuscript. All authors contributed to the discussion of the results and the manuscript.

\smallskip{}

\noindent \textbf{\large{}Competing interests}{\large\par}

\noindent The authors declare no competing interests.

\end{document}



\onecolumngrid
\pagebreak
\widetext
\begin{center}
\textbf{ {\large - Supplemental Information -} \\
{\large Generalized Number-Phase Lattice Encoding of a Bosonic Mode for Quantum Error Correction}}
\end{center}

\setcounter{equation}{0}
\setcounter{figure}{0}
\setcounter{table}{0}

\renewcommand{\theequation}{\arabic{equation}}
\renewcommand{\thefigure}{\arabic{figure}}
\renewcommand{\thetable}{\arabic{table}}
\renewcommand{\figurename}{\textbf{Supplementary Figure}}
\renewcommand{\tablename}{\textbf{Supplementary Table}}

\noindent \textbf{\large{}Supplementary Notes 1- Number-phase representation of a bosonic mode.}
\smallskip
\smallskip
\smallskip
\smallskip

In the main text, we have shown that an operator performed on a bosonic mode can be decomposed by the number-phase (NP) displacement operators basis. Here we supplement the component for the NP representation of the quantum states. In general, a quantum state can be expanded using the Fock basis as $|\psi\rangle=\sum_nc_n|n\rangle$, while it also can be alternatively unfolded using the Pegg-Barnett phase state $|\phi\rangle=(2\pi)^{-1/2}\sum_{n=0}^{\infty} e^{in\phi}|n\rangle$. The phase states are the Fourier transformation of the Fock states, which satisfy the relationship of orthogonality $\langle\phi|\phi'\rangle=\delta(\phi-\phi')$ and the normalization $\int_0^{2\pi}d\phi|\phi\rangle\langle\phi|=\hat{I}$. In addition, the phase states are eigenstates of the operator $\hat{\Sigma}_l$ that is $\hat{\Sigma}_l|\phi\rangle=e^{il\phi}|\phi\rangle$. By using the normalization of phase states, we can obtain the orthogonality between NP displacement operators
\begin{align}
        \mathrm{Tr}(\hat{\mathcal{D}}^\dagger(\mathbf{n}')\hat{\mathcal{D}}(\mathbf{n}))&=\int_0^{2\pi}dx\langle x|(\hat{\mathcal{D}}^\dagger(\mathbf{n}')\hat{\mathcal{D}}(\mathbf{n})|x\rangle\\\nonumber
        &=e^{\frac{i\phi l+i\phi'l'-i2\phi l'}{2}}\int_0^{2\pi}dx \langle x |(\hat{R}(\phi-\phi')\hat{\Sigma}_{l-l'}|x\rangle\\\nonumber
        &=e^{\frac{i\phi (l-l')}{2}}\delta(\phi-\phi')\int_0^{2\pi}dx e^{i(l-l')x}\\\nonumber
        &=2\pi\delta(l-l')\delta(\phi-\phi')\\\nonumber
        &=2\pi\delta^2(\mathbf{n}-\mathbf{n}')\\\nonumber
\end{align}

Moreover, to make the codewords of generalized NP codes to be visual in the phase space defined by number and phase variables, we introduce the number-phase Wigner function~\cite{Vaccaro1995} representation form. It is a quasiprobability distribution of quantum states defined by number variable $\hat{n}$ and phase variable $\hat{\phi}$, instead of the original Wigner function is defined by position and momentum variables. In practice, the NP Wigner function can be defined in Fock basis\cite{Vaccaro1995}
\begin{align}
        &W_{NP}(n,\phi)={\rm Tr}(\rho\hat{S}_{NP}(n,\phi)),\\
    &\hat{S}_{NP}(n,\phi)=\frac{1}{2\pi}(\sum_{p=-n}^{n}e^{2ip\phi}|n+p\rangle\langle n-p|+\sum_{p=-n}^{n-1}e^{i(2p+1)\phi}|n+p\rangle\langle n-p-1|).
    \label{S2}
\end{align}
The second term of Eq.~(\ref{S2}) is conventionally set to 0 when p=n. Using the NP Wigner function representation of NP codes, one can intuitively observe the the number-phase lattice structure, as shown in Supplementary Figure~1. This physical image is similar to that of GKP codes, and one can say that both are lattices in phase space, while the difference is that the number variable is discrete and the phase variable is periodic. If the lattice exhibits translational symmetry along the hybridized number-phase variable direction, the periodicity of the phase variable will lead to a vortex structure in the number-phase representation, such as the oblique and diamond lattice shown as Fig.~1 (e) in the main text.

\begin{figure}[thb]	
\centering
\includegraphics[width=0.75\linewidth]{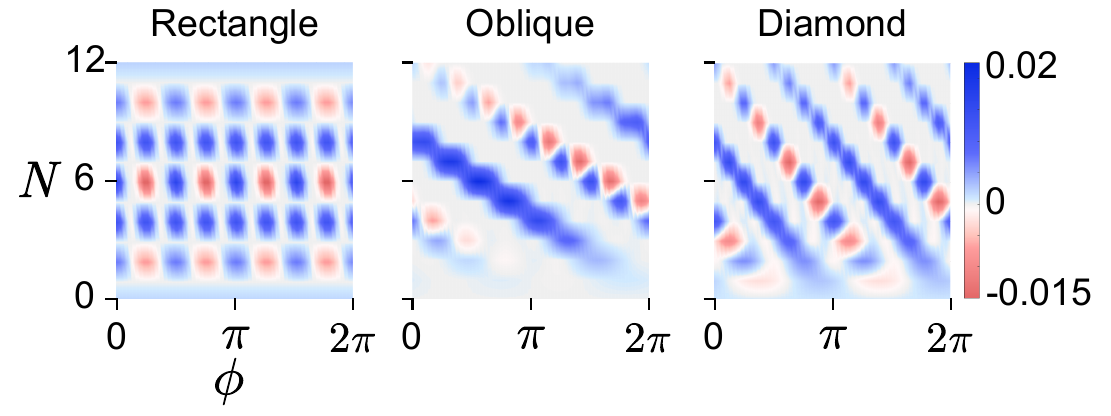}
\caption{NP Wigner distributions of three generalized NP codes defined in Fig.~2 of the main text. (a) the binomial code, (b) the O-NP code, and (c) the D-NP code. }
\label{Fig5}
\end{figure}

\noindent \textbf{\large{}Supplementary Notes 2- Generalized NP codes with finite energy.}
\smallskip
\smallskip
\smallskip
\smallskip

In the main text, we provide the NP Wigner function representation of the codespaces of the ideal generalized NP codes, here we use this representation to prove that the correctable error set $\bar{E}=\{\mathcal{D}(\mathbf{n}_e)\}$ with $\mathbf{n}_e\in\{~(l_e,\phi_{e})~|-G\leq l_e\leq L,~~|\phi_{e}|<\frac{\pi}{2d_N} \}$ exactly satisfy the KL conditions 
\begin{align}
    \langle\mu|\hat{\mathcal{D}}^\dagger(\mathbf{n}_e')\hat{\mathcal{D}}(\mathbf{n}_e)|\nu\rangle_L=h_{\mathbf{n}_e,\mathbf{n}_e'}\delta_{\mu,\nu},
    \label{S4}
\end{align}
where $\mu,\nu$ run the logical $0,1$. The error in $\bar{E}$ can be entirely removed by a error-correcting code when the KL conditions are exactly satisfied. Due to the NP Wigner function of the codespaces of ideal generalized NP codes is lattice in NP phase space, the non-zero value in Eq.~(\ref{S4}) is only present when the NP displacement error $\hat{\mathcal{D}}^\dagger(\mathbf{n}_e')\hat{\mathcal{D}}(\mathbf{n}_e)$ simultaneously commutes with stabilizers $\hat{S}_{X}=\hat{\mathcal{D}}(2\mathbf{n}_x)$ and $\hat{S}_{Z}=\hat{\mathcal{D}}(2\mathbf{n}_z)$. In other words, only when the NP displacement error accumulates to a level that causes logical errors does it fail to satisfy the KL conditions. Therefore, the displacement error $\hat{\mathcal{D}}^\dagger(\mathbf{n}_e')\hat{\mathcal{D}}(\mathbf{n}_e)$ with $|l_e-l_e'|\leq G+L=d_N-1$ and $|\phi_e-\phi_e'|< \pi /d_N$ all do not commute with the two stabilizers, thus it is exactly satisfying the KL conditions Eq.~(\ref{S4}) with zero values, expect for the identity error $\hat{I}$ which obviously satisfy the KL conditions and have a non-zero value. Similarly, the error set $\mathbf{n}_e\in\{~(0,\phi_{e})~|~|\phi_{e}|<{d_{\phi}}/{2} \}$ with $|l_e-l_e'|=0$ and $|\phi_e-\phi_e'|< d_\phi$ also is a correctable error set.

In experiments, the ideal generalized NP codes is not physical, thus we give the definition of generalized NP codes [Eq.~(8) in the main text] by introducing the normalizable Fock-amplitude $\{\theta_n\}$. This Fock-amplitude not only gives the mean excitation of the generalized NP code $\bar{n}_{\mathrm{code}}=s\sum_{n=0}^{\infty} n|\theta_n|^2$, but also gives the NP variable uncertainty $\Delta_{\phi}=\sum_n|\theta_n\theta_{n-1}|^{-2}-1$ of the codewords in $\mathbf{n}_x$-direction, by using the phase measurement $\mathcal{U}_f(\hat{\mathcal{M}}_X)=U_f\mathcal{M}_X U_f^\dagger$ and $\hat{\mathcal{M}}_X$ is the canonical phase measurement. The non-zero uncertainty of NP variables results in that the KL conditions are no longer strictly satisfied for generalized NP codes, since an arbitrarily small phase rotation can lead to non-zero phase-flip $\langle-|\hat{R}(\delta\phi)|+\rangle_L\neq0$. For the situation $\theta_n=\langle n|\alpha,r\rangle$ and $\theta_n=\sqrt{2^{-K}\binom{K}{n}}$ which we are concerned in the main text, the mean excitation of NP codes are $s[|\alpha|^2+\sinh^2(r)]$ and $Ks/2$, respectively. To quantify the imperfection of generalized NP codes for correcting the error set $\bar{E}$, we introduce the KL cost function 
\begin{align}
    C_{KL}(\bar{E})=\int d\mathbf{n}_ed\mathbf{n}_e'\{|\langle 0|\hat{\mathcal{D}}^\dagger(\mathbf{n}_e')\hat{\mathcal{D}}(\mathbf{n}_e)|1\rangle_L|^2+|\langle 0|\hat{\mathcal{D}}^\dagger(\mathbf{n}_e')\hat{\mathcal{D}}(\mathbf{n}_e)|0\rangle_L-\langle 1|\hat{\mathcal{D}}^\dagger(\mathbf{n}_e')\hat{\mathcal{D}}(\mathbf{n}_e)|1\rangle_L|^2\}.
\end{align}
The smaller KL cost function indicates a better error correction capability for a given error set $\bar{E}$, and the vanished KL cost function signifies the exact error correction. In Supplementary Figure~\ref{Fig.S2} (a), we numerically demonstrate that the KL cost function for the above tow error sets both exponentially decreases as mean code excitation of generalized NP codes increases, where we select the squeezing parameter $r=0$ for an example.
\begin{figure}[thb]	
\centering
\includegraphics[width=1\linewidth]{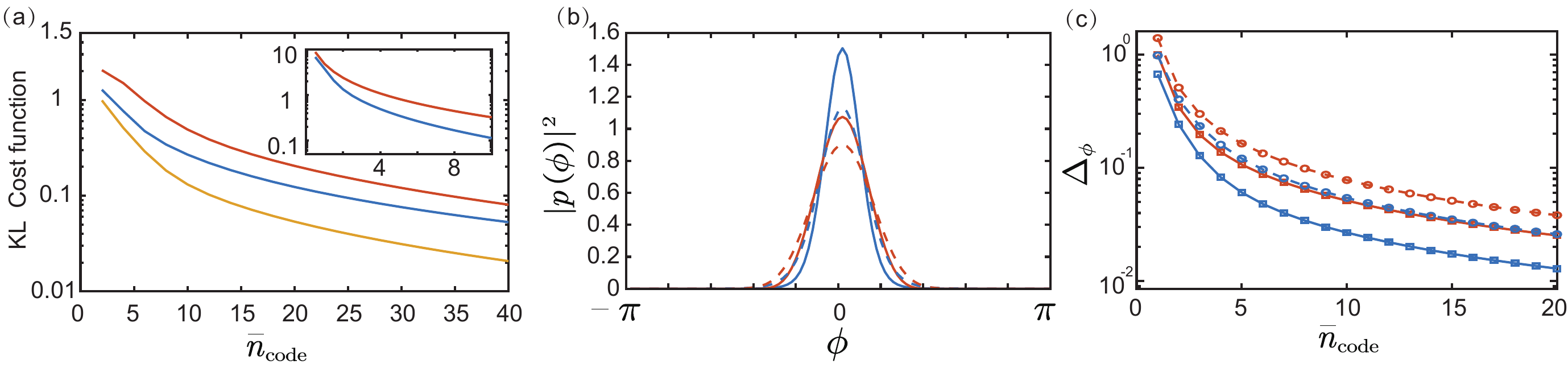}
\caption{Nonzero KL cost function and phase uncertainty of generalized NP codes with finite energy. (a) The KL cost function $C_{\mathrm{KL}}(\bar{E})$ of generalized NP codes varies with the mean code excitation $\bar{n}_{\mathrm{code}}$, and they have a same code distance $d_N=4$. The insert is analogous, except that the error set has been altered as $\mathbf{n}_e\in\{~(0,\phi_{e})~|~|\phi_{e}|<{d_{\phi}}/{2} \}$ for a same code distance $d_{\phi}=\pi$. (b) The probability distribution of the logical basis $|+\rangle_L$ in $\mathbf{n}_x$-direction with different $\theta_n$ and phase measurement, and (c) illustrates phase uncertainty of this logical basis varies as functions of $\bar{n}_{\mathrm{code}}$. Here, orange, blue, and yellow lines represent binomial, O-NP, and D-NP codes, respectively. The solid (dashed) lines indicate the results obtained by canonical phase (heterodyne) measurement.}
\label{Fig.S2}
\end{figure}
In addition, the phase measurement is the key component to measure the error syndrome for phase rotation, and the phase uncertainty not only depends on the codewords, but also depends on the phase measurement. For a given codeword, the different phase measurements can give different phase uncertainty, and a larger phase uncertainty indicates more measurement error. The general phase uncertainty $\Delta_{\phi}$ of the codeword $|+\rangle_L$ with a phase rotation symmetry $\hat{R}(2\pi/s)$ can be defined as~\cite{Holevo2011} 
\begin{align}
    \Delta_{\phi}=|\langle\mu(\phi)e^{is\phi}\rangle|^{-2}-1,
\end{align}
where $\mu(\phi)=\mathrm{Tr}(\hat{M_{\phi}}|+\rangle\langle+|_L)$ is the general phase probability distribution of the codeword $|+\rangle_L$ given by the alternative phase measurement $\hat{M}_{\phi}$, and $\langle\mu(\phi )e^{is\phi}\rangle=\int_0^{2\pi}d\phi~\mu(\phi )e^{is\phi}$ is the mean modular phase. There are two state-irrelevant phase measurements, the canonical phase measurement and the heterodyne measurement whose positive operator-valued measure (POVM) element are $\hat{\mathcal{M}}_{\phi}^C=|\phi\rangle\langle\phi|$ and $\hat{\mathcal{M}}_{\phi}^H=\pi^{-1}\int_0^{\infty}d\alpha~|\alpha e^{i\phi}\rangle\langle\alpha e^{i\phi}|\alpha$, respectively. The two phase measurements both satisfy the identity relation $\int_{0}^{2\pi}d\phi\hat{\mathcal{M}}_{\phi}=\hat{I}$. In terms of the phase measurement accuracy, the canonical phase measurement always performs better than the heterodyne measurement, since the heterodyne measurement projects a quantum state onto a coherent state which not only gives the estimated phase, but also gives the coherent amplitude which is wasted. However, the heterodyne measurement is a well-known Gaussian POVMs whose experimental realization is technical mature, while the implementation of the canonical phase measurement with high Hilbert space still face practical challenges. In Supplementary Figures~\ref{Fig.S2} (b) and (c), we numerically compare the phase probability distribution and the phase uncertainty of the codeword $|+\rangle_L$ by using the different phase measurements, which confirms the advantages of the canonical phase measurement. These tow figures are obtained by selecting $s=1$ and the only difference of the situation with $s>1$ is that $\bar{n}_{\mathrm{code}}$ need be replaced as $s\bar{n}_{\mathrm{code}}$ for a same phase uncertainty value.   

\smallskip
\smallskip
\smallskip
\smallskip

\noindent \textbf{\large{}Supplementary Notes 3- Interface gate and readout of phase rotation.}
\smallskip
\smallskip
\smallskip
\smallskip

In this section, we introduce the interface gate that allows mutual conversion between two different types of generalized NP codes. Then we graphically show the interface between R-NP codes and O-NP codes, and the NP vortex effect in NP space.
\begin{figure}[thb]	
\centering
\includegraphics[width=0.7\linewidth]{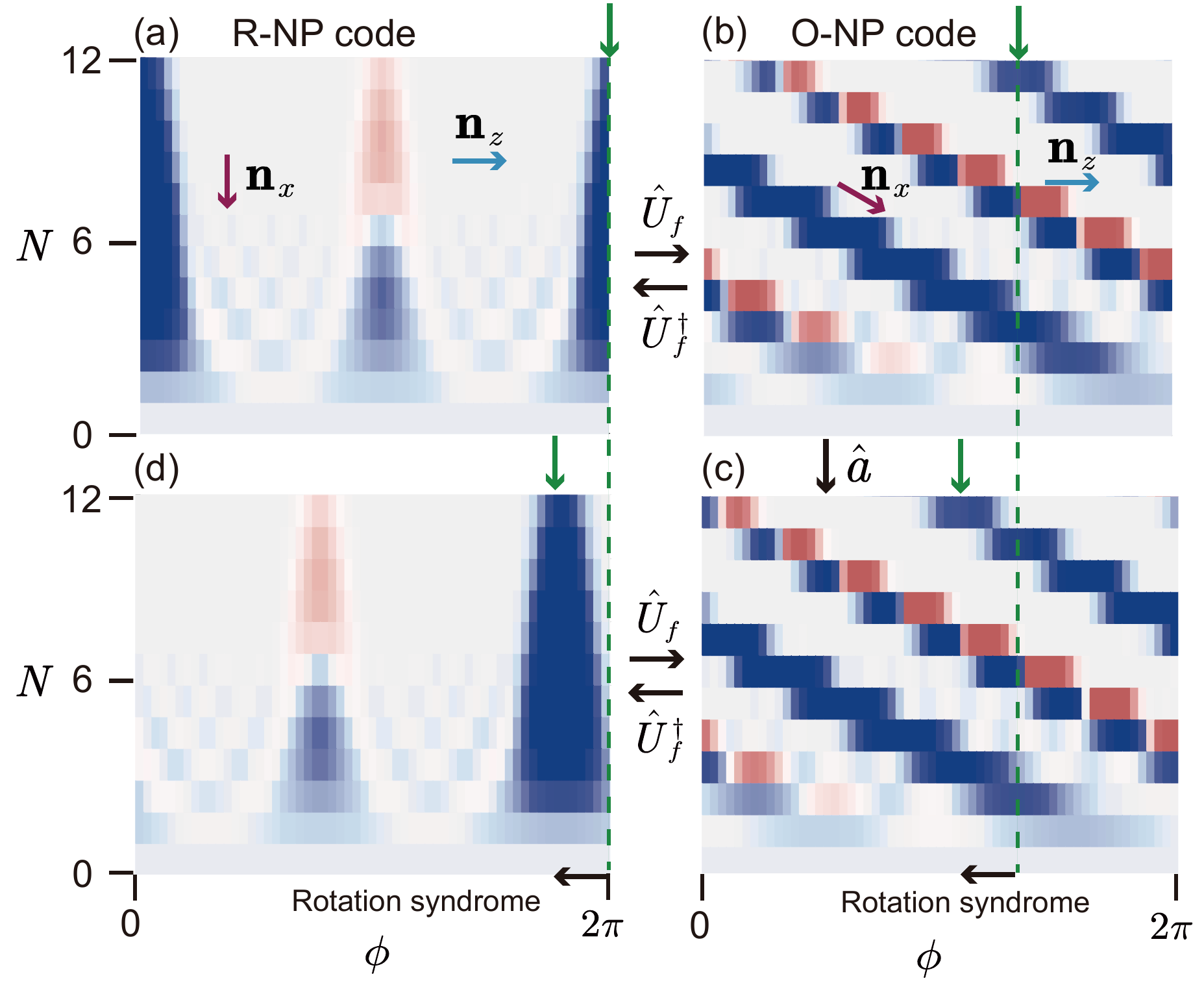}
\caption{Graphical illustration of codespace transformation of generalized NP codes via the interface gate $\hat{U}_{f=1/4}$. The NP Wigner functions are presented for codeword $|+\rangle_L$ of (a) the R-NP code  $[s=1;f=0;r=0]$ and (b) the O-NP code $[s=1;f=1/4;r=0]$, with the same mean code excitation $\bar{n}_{\mathrm{code}}=6$. (c) shows the result of single-photon loss applied to (b), while (d) illustrates the state after applying the inverse interface gate $\hat{U}^\dagger_{f=1/4}$ to (c). } 
\label{FigS3}
\end{figure}

The interface gate can achieve the transformation between NP lattice structure $(s,f_1)\leftrightarrow(s,f_2)$, which is mathematically defined as
\begin{equation}
  \hat{U}_s(\Delta f)\coloneqq \exp(-\frac{i\Delta f\pi\hat{n}^2}{2s^2}),
\end{equation}
where $\Delta f=f_2-f_1$. Note that the interface gate changes the codespaces via only modifying the Pauli X operator
\begin{equation}
  \hat{U}_s(\Delta f)\hat{\mathcal{D}}(\mathbf{n}_x)\hat{U}^\dagger_s(\Delta f)=\hat{\mathcal{D}}[\mathbf{n}_x+(0,\frac{\Delta f\pi}{s})],
\end{equation}
without affecting the Pauli Z operator $\hat{\mathcal{D}}(\mathbf{n}_z)$ that depends the rotation symmetry of the codespaces. As we showed in the main text, the interface gate can be used to change the code distance $d_N$ of the generalized NP codes. For example, we graphically illustrate the transformation of codewords $|+\rangle_L$ between the R-NP code $[s=1,f=0,r=0]$ and the O-NP code [Supplementary Figure~\ref{Fig5}(b)] , with a shared mean code excitation $\bar{n}_{\mathrm{code}}=6$, as shown in Supplementary Figure~\ref{FigS3}(a) and (b). Here the code distance $d_N=1$ of the R-NP code is unitarily transformed to be $d_N=4$ for the O-NP code through the interface gate $\hat{U}_{f=1/4}$, while the code distance $d_{\phi}=\pi$ remains unchanged. 

Another main use of such a gate is to interface the generalized NP codes $(s,f)$ and the R-NP codes $(s,0)$ with the shared parameter $s$. Due to the qubit operations $\mathcal{P}_s$ for R-NP codes have been widely studied in experiments, the interface gates allow these operations to be compatible with the generalized NP codes as $\mathcal{P}_{s,f}=\hat{U}_s(f)\mathcal{P}_s\hat{U}_s^\dagger(f)$. This can be understood as a two-step process: first, transferring a generalized NP code $(s,f)$ to R-NP code $(s,0)$ to execute qubit operations, and then transferring it back to the original code spaces to complete the qubit operations for the generalized NP code $(s,f)$.

Given that any phase rotation commutes with interface gates, a critical application of these gates is to map a noisy generalized NP code onto the R-NP structure. This process enables the discrete phase rotation—induced by the NP vortex effect—to be readout using canonical phase measurements. The graphical representation of this process is shown in Supplementary Figure~\ref{FigS3},  where we demonstrate the discrete phase rotation caused by single-photon loss. Before using the interface gate $\hat{U}_f^\dagger$, the phase rotation of generalized NP codes [Supplementary Figure~\ref{FigS3}(c)] can not be directly readout by the canonical phase measurement. This is because the codeword is only localized in the direction orthogonal to $\mathbf{n}_x$, rendering its canonical phase distribution non-local. After performing the interface gate $\hat{U}_f^\dagger$ [Supplementary Figure~\ref{FigS3}(d)], $\mathbf{n}_x$ becomes orthogonal to the canonical phase, localizing the codeword's distribution in the canonical phase. Consequently, the phase rotation ($\Delta\phi=\pi/4$) can be accurately readout by the canonical phase measurement. 

\smallskip
\smallskip
\smallskip
\smallskip

\noindent \textbf{\large{}Supplementary Notes 4- Quantum error correction with generalized NP codes.}
\smallskip
\smallskip
\smallskip
\smallskip

Different from the well-known R-NP codes that uses error subspaces with different number parity to avoid number-shift errors, the generalized NP codes suggest that the rotated codespaces also serves as error subspaces for the number shift errors based on the practical number-phase vortex structure. Here, we consider the O-NP and the D-NP code as examples to demonstrate the error subspaces distribution in the continuous-variable phase space after performing the interface gate $\hat{U}^\dagger_f$, as shown in Supplementary Figure~\ref{Fig.S3}. These rotated codewords can be distinguished by the phase measurement mentioned in the above section, and the measurement outcomes give error syndromes. 

In the main text, we proposed the recovery circuit of the noisy O-NP codes based on an one-step teleportation. The main mechanism is the controlled phase gate induced bell entanglement between the data mode and the ancilla mode.
\begin{align}
    \bar{C}_Z\hat{\mathcal{D}}(l_e,\phi_e)|\psi\rangle_L\otimes|+\rangle_L=e^{i\Phi}\sum_{i=0,1}\hat{R}(-l_ef\pi+\phi_e)|\tilde{(-)^i}_{l_e}\rangle_L\otimes\bar{X}^i\bar{Z}^{l_e}\bar{H}|\psi\rangle_L.
\end{align}
Here $e^{i\Phi}$ is an unimportant global phase, and $|\phi_e|<\pi/(2q)$ is treated as a distinguishable disturbance of rotated codewords, and $|\tilde{\pm}_{l_e}\rangle_L$ is the number-shifted codewords. A further error syndromes measurement need correctly identify the rotated codewords with the presence of noise and simultaneously teleports the quantum state from the data mode to the ancilla mode with additional logical operations. According to the outcome of error syndromes, the logical information can be recovery through the feedback $\hat{\mathcal{R}}=\bar{H}\bar{Z}^{l_e}\bar{X}^i$. 

Similarly, the recovery of noisy generalized NP codes also based on the teleportation circuit, but it requires to implement a number parity measurement $\hat{P}_{s,k}=\sum_{n=1}^{\infty}|ns-k\rangle\langle ns-k|$ before the teleportation, where the integer $k\in[0,s-1]$ represent the general parity. The realization of such a parity measurement is varied, and we proposed a type of the modular number parity measurement in the main text. A noisy generalized NP code after performing the number parity measurement becomes   
\begin{align}
    \hat{P}_{s,k}\hat{\mathcal{D}}(l_e,\phi_e)|\psi\rangle_L=e^{i(ms+k)\phi_e/2}\hat{R}(\phi_e)\Sigma_{ms+k}|\psi\rangle_L.
\end{align}
Namely, the number-shit with $k=l_e~\mathrm{mod}~s$ is fixed by the number parity measurement, and the only variable of $l_e$ is $m$. Then the noisy generalized NP code requires to be sent to the teleportation-based QEC circuit. Since $k$ is fixed, the every $s$ number-shift will lead to a $\bar{X}$ logical error and also result in a rotation $\hat{R}(-mf\pi/s)$ of codewords due to the number-phase vortex effect, and the maximal detectable $m$ is $q-1$ which is similar to the O-NP code. In addition, the number-phase vortex effect will result in a global rotation $\hat{R}(-kf\pi/s^2)$ of the data mode due to the fixed $k$, and one need to get the error syndromes $m$ and $i$ from the outcome of the phase measurement $X=(\pi/s)^i-kf\pi/s^2-mf\pi/s+\phi_e$ in the teleportation, and the detectable $l_e=ms+k<qs=d_N$. Moreover, the controlled phase gate $\bar{C}_Z$ will lead to a phase rotation $\hat{R}(k\pi/s)$ of the ancilla mode as the reaction, and this phase rotation can be written as $\bar{Z}^k$, since $\hat{R}(\pi/s)$ is the logical $\bar{Z}$ operation of the generalized NP codes, and thus the recovery operation $\hat{\mathcal{R}}=\bar{H}\bar{Z}^m\bar{X}^i\bar{Z}^k$.

\begin{figure}[t]	
\centering
\includegraphics[width=0.9\linewidth]{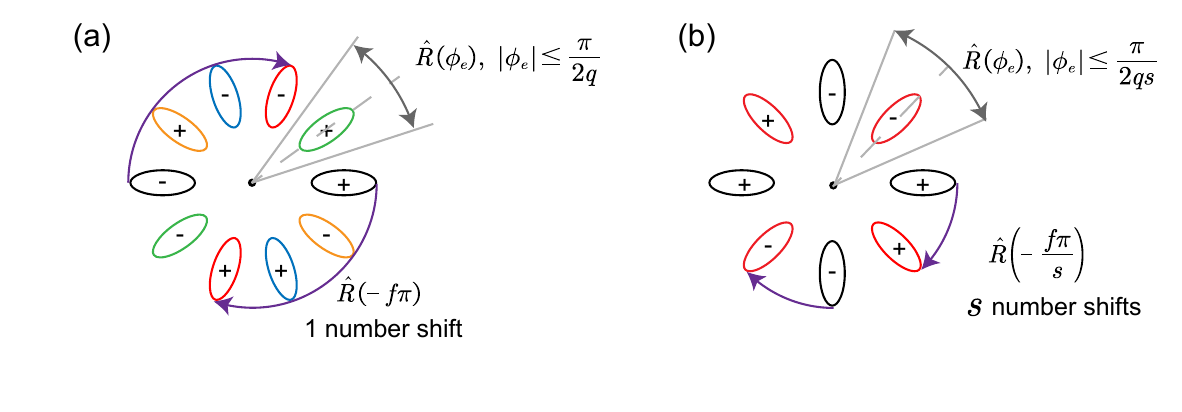}
\caption{Error spaces transformation of generalized NP codes in quadrature phase space. (a) The sketch for error subspaces of O-NP codes in continuous-variables phase space, after performing the interface gate $\hat{U}_f$. Here, the parameter $f=p/q=3/5$ and black ovals represent the original codespaces, while different colored ovals indicate different error subspaces. (b) The sketch for error subspaces of D-NP codes after performing the interface gate $\hat{U}_f$, and the code parameters are selected as $s=2,~f=1/2$. The marks `$+$' and `$-$' represent the logical basis $|+\rangle_L$ and $|-\rangle_L$, respectively.}
\label{Fig.S3}
\end{figure}

As a practical example, we considered the error model $\dot{\hat{\rho}}=\gamma\mathcal{L}[\hat{a}]\hat{\rho}+\kappa\mathcal{L}[\hat{a}]\hat{\rho}$ in the main text, which gives the noise channel $\mathcal{N}(\hat{\rho})=e^{\gamma t\mathcal{L}[\hat{a}]+\kappa t\mathcal{L}[\hat{n}]}\hat{\rho}$. Here we give the Kraus-operator decomposition of the noise channel $\mathcal{N}(\hat{\rho})$ as
\begin{align}
   \mathcal{N}[\hat{\rho}]=\int_{-\infty}^{\infty} d\phi P(\phi) \sum_{l=0}^{\infty}\hat{R}(\phi)\hat{A}_l\hat{\rho} \hat{A}_l^\dagger \hat{R}(\phi)^\dagger,
   \label{S9}
\end{align}
which also can be found in the Ref.~\cite{Grimsmo2020}. Here $\hat{A}_l=(1/\sqrt{l!})\Gamma^{j/2}(1-\Gamma)^{\hat{n}/2}\hat{a}^l$ and $\Gamma=1-e^{-\gamma t}$ is the excitation loss rate and $P(\phi)=(1/\sqrt{2\pi\kappa t})\exp(-\phi^2/2\kappa t)$ is a Gaussian distribution with the variance $\kappa t$. Therefore, we use  $\hat{N}_j=\sqrt{P(\phi)}\hat{R}(\phi)\hat{A}_l$ to refer the Kraus-operator decomposition of the noise channel $\mathcal{N}$. Furthermore, we illustrate that the QEC processes of O-NP codes with the code states passed through the noise channel $\mathcal{N}(\rho)$ by using NP phase space representation, shown in Supplementary Figure~\ref{Fig. S4}. Here, the O-NP code allows the number-shift errors to be equivalent to phase rotation errors, and the phase probability distribution of the noisy O-NP code after performing the interface gate $\hat{U}_f^\dagger$ is shown in the Supplementary Figure~\ref{Fig. S4} (b2). Based on the NP vortex effect, the peaks of the phase probability distribution represents the number-shifts induced phase rotation, and the higher-order number-shift will result in a lower peak due to the less probability of occurrence. Therefore, one can perform the feedback operation on the teleported ancilla mode according to the outcome $X$ of the canonical phase measurement. 

Different from the number-parity, the rotated codewords are not exactly orthogonal due to the non-vanished phase uncertainty. The key problem is how to manage the intervals of the measurement outcomes, thereby correctly implementing the recovery operations for achieving the maximal channel fidelity. This problem is difficult to solve analytically, but it can be easily solved by a simple numerical optimization. For a given noise channel $\mathcal{N}$, generalized NP codes $|\pm\rangle_L$, and the phase measurement $\hat{\mathcal{M}}_{X}$, the recovery operation of the teleported ancilla mode only have four different situations: $\hat{\mathcal{R}}_{h,t}=\bar{H}\bar{Z}^h\bar{X}^t\bar{Z}^k$ with $h=0,1$ and $t=0,1$. For an outcome $X$ of the phase measurement, the classical computer can give a best recovery operation among the four situations which gives the maximal fidelity. Therefore, every outcome $X$ corresponds to a best recovery operation, thus we can manage the intervals of the outcomes according to the priori best recovery operation given by the classical computer. For an example in Supplementary Figure~\ref{Fig. S4} (b2), the every outcome in intervals $(x_i,x_{i+1})$ corresponds to a same best recovery operation, and the adjacent tow intervals have two different best recovery operations with the separation node $x_i$.  

\begin{figure}[tbp]	
\centering
\includegraphics[width=0.75\linewidth]{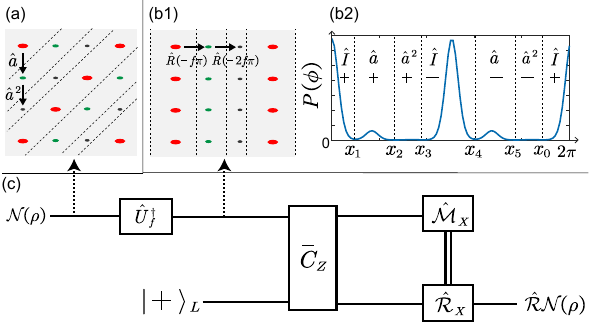}
\caption{QEC implementation of the O-NP code with $f=-1/3$. (a) Schematic diagram of the NP Wigner function for noisy O-NP codes $\mathcal{N}(\rho)$, where the every arrow represents a photon loss event. (b1) shows that after performing the interface gate to (a), where the every arrow represents a discrete phase rotation, and (b2) is the corresponding phase probability distribution obtained by the canonical phase measurement. The red ovals represent the original code subspace, and the blue and grey ovals represent the error subspaces, respectively.  (c) Circuit schematic of the implementation of O-NP codes. }
\label{Fig. S4}
\end{figure}

\smallskip
\smallskip
\smallskip
\smallskip

\noindent \textbf{\large{}Supplementary Notes 5- QEC performance of generalized NP codes.}
\smallskip
\smallskip
\smallskip
\smallskip

The QEC performance of generalized NP codes with small mean code excitation $\bar{n}_{\mathrm{code}}$ usually is different due to the large phase uncertainty and various choices of the Fock-amplitude $\{\theta_n\}$, however, they have a similar behavior when the phase uncertainty is small enough ($\bar{n}_{\mathrm{code}}$ is large). Here we summarize the similar behaviors of generalized NP codes when the phase uncertainty of the codewords are small. The following discussion is based on the near-optimal performance of the generalized NP codes~\cite{Zheng2024}, which is the channel fidelity of the near-optimal transpose channel~\cite{Barnum2002} and is defined as $\tilde{\mathcal{F}}=||\mathrm{Tr}_L\sqrt{M}||_F^2/4$, where $M$ is the QEC matrix and $||\cdot||_F$ is the Frobenius norm. The QEC matrix $M$ is defined as $M_{[\mu,j][\nu,k]}=\langle\mu|\hat{N}^\dagger_j\hat{N}_k|\nu\rangle_L$, and $\mathrm{Tr}_L(\cdot)$ indicates to trace the logical index $\mu$ and $\nu$. Moreover, the near-optimal infidelity can be obtained as $1-\tilde{\mathcal{F}}\approx||g(D)\odot\Delta M_||^2_F/2$, where $D=\mathrm{Tr}_L(M)/2$ is a diagonal matrix, $\Delta M=M-\sigma_I\otimes D$, and $g(D)_{[\mu j],[\nu k]}=1/(\sqrt{d_j}+\sqrt{d_k})$. We use $\sigma_i$ with $i\in\{I,x,y,z\}$ to refer the corresponding $2\times2$ Pauli matrix, and $d_j$ is the $j$-th eigenvalue of the diagonal matrix $D$ and $\odot$ is the Hadamard product.

Let us consider the noise channel Eq.(~\ref{S9}) and a given generalized NP codes defined by parameters $s$ and $f$. In the limitation that the phase uncertainty of codewords is small enough and the mean code excitation is large, the dominated error is bit-flip error and the phase-flip error is negligible, and the Fock amplitude of generalized NP codes becomes consistent with $\theta_n\approx\langle sn|\alpha\rangle$. Because the binomial distribution approaches the Poisson distribution when $n$ is large enough, and the squeezing effect of displaced squeezed state $|\alpha,r\rangle$ also can be ignored when $|\alpha|^2\gg \sinh^2(r)$. Under such a approximation, the matrix $D=\mathrm{Tr}_L(M)/2$ is diagonal with the eigenvalue $d_j=(\Gamma\bar{n}_{\mathrm{code}})^je^{-\Gamma\bar{n}_{\mathrm{code}}}/l!$. In addition, $\Delta M=\sigma_x\otimes M^x$ with $M^x_{j,k}=(\delta_{j,0}\delta_{k,ld_N}+\delta_{k,0}\delta_{j,ld_N})(\bar{n}_{\mathrm{code}}\Gamma)^{(j+k)/2}(j!k!)^{-1/2}e^{-\Gamma\bar{n}_{\mathrm{code}}}$ with $l\in\mathbb{Z}^+$ describe the dominant phase-flip error. Therefore, the near-optimal infidelity of generalized NP codes in the large-excitation limitation can be estimated as 
\begin{align}
   1-\tilde{\mathcal{F}}^{\gamma}&=\frac{1}{2}||g(D)\odot(\sigma_x\otimes M_x)||^2_F\\\nonumber
   &=\sum_{j,k}\frac{|M^x_{j,k}|^2}{(\sqrt{d_j}+\sqrt{d_k})^2}\\\nonumber &\approx2\sum_{l=1}\frac{(\bar{n}_{\mathrm{code}}\Gamma)^{ld_N}e^{-2\Gamma\bar{n}_{\mathrm{code}}}}{d_0(ld_N)!}\\
   &=2\sum_{l=1}\frac{(\bar{n}_{\mathrm{code}}\Gamma)^{ld_N}e^{-\Gamma\bar{n}_{\mathrm{code}}}}{(ld_N)!}.\nonumber
\end{align}
The third line omits the high-order small quantity $d_{ld_N}$, and this estimation of the performance for generalized NP codes is valid when the $\Gamma\bar{n}_{\mathrm{code}}\ll1$. They main result is that the performance of the generalized NP codes with a large $\bar{n}_{\mathrm{code}}$ depends on the code distance $d_N$ and the enhanced loss rate $\Gamma\bar{n}_{\mathrm{code}}$.

Moreover, if the excitation loss rate $\Gamma$ in Eq.~(\ref{S9}) is vanished, i.e., there are only phase rotation errors, the generalized NP codes can monotonically suppress the phase rotation errors as $\bar{n}_{\mathrm{code}}$ increasing. In the limitation that $\bar{n}_{\mathrm{code}}$ of generalized NP codes is infinite, we can give the lower bound of performance for correcting the phase rotation errors. Since the phase rotation error is continuous, the QEC matrix $M$ is infinite dimension and the diagonal matrix $D_{\phi,\phi}=P(\phi)$. In Eq.~(\ref{S9}), the phase $\phi\in(-\infty,\infty)$, and the value range can be reduced to $(-\pi,\pi)$ by using the periodicity of the phase variable, and the probability distribution of the phase errors simultaneously becomes $P_c(\phi)=(2\pi)^{-1}\vartheta_3(\phi/2,e^{-\kappa t/2})$, by using the Poisson summation formula. Here, $\vartheta_3$ is the third kind of the elliptic theta function, which is defined as $\vartheta_3(x,a)=1+2\sum_{n=1}^{\infty}a^{n^2}\cos(2nx)$. The phase-flip error can be described by $\Delta M=\sigma_z\otimes M^z$ and $M^z_{\phi,\phi'}=(\delta_{\phi,\phi'+d_{\phi}}+\delta_{\phi,\phi'-d_{\phi}})\sqrt{P_c(\phi)P_c(\phi')}$, thus the lower bound can be obtained as
\begin{align}
   (1-\tilde{\mathcal{F}}^{\kappa})_{\mathrm{min}}&=\frac{1}{2}||g(D)\odot(\sigma_z\otimes M_z)||^2_F\\
   \label{S12}
   &=\int_{-\pi}^{\pi} d\phi\frac{2P_c(\phi-d_{\phi})P_c(\phi)}{(\sqrt{P_c(\phi-d_{\phi})}+\sqrt{P_c(\phi)})^2}\\ 
   \label{S13}
   &\approx\frac{1}{2}e^{-\frac{d_{\phi}^2}{8\kappa t}}\int_{-\infty}^{\infty} d\phi P_c(\phi)\mathrm{sech}^2(\frac{\phi d_{\phi}}{4\kappa t})\\
   \label{S14}
   &\approx e^{-\frac{d_{\phi}^2}{8\kappa t}}P(0)\frac{4\kappa t}{d_{\phi}}\\
   &=\frac{(8\kappa t/d_{\phi}^2)^{\frac{1}{2}}}{\sqrt{\pi}}e^{-\frac{d_{\phi}^2}{8\kappa t}}.
\end{align}
One can obtain Eq.~(16) in the main text, by explicitly expressing $d_{\phi}=\pi/s$. Eq.~(\ref{S12}) is the exact form of the lower bound, but it is not easy to understand. Therefore, we give the approximate form of the lower bound when the assumption $\kappa t\ll1$ is valid. Such an assumption indicates that the probability distribution of the phase rotation error $P(\phi)$ concentrates on $\phi=0$ and which is negligible in the range $\phi\in(-\infty,-\pi)\cup(\pi,\infty)$. Therefore, we have the approximations $P_c(\phi)\approx P(\phi)$ and $\int_{-\pi}^{\pi}\cdot ~d\phi\approx \int_{-\infty}^{\infty}\cdot ~d\phi$, which gives the Eq.~(\ref{S13}). Furthermore, by using the integral $\int_{-\infty}^{\infty}d\phi \cos(n\phi)\mathrm{sech}^2(c\phi)=n\pi\mathrm{csch}[n\pi/(2c)]/c^2\approx2/c$ when $c\ll1$, we have the approximation $\int_{-\infty}^{\infty} d\phi P_c(\phi)\mathrm{sech}^2(\frac{\phi d_{\phi}}{4\kappa t})\approx\frac{ 8\kappa t}{d_{\phi}}P_c(0)\approx\frac{ 8\kappa t}{d_{\phi}}P(0)$, which gives the Eq.~(\ref{S14}). The approximate form of the lower bound for the generalized NP codes reveals that the pure rotation error rate is quadratically increasing as the code distance $d_{\phi}$ linearly decreasing. 

\smallskip
\smallskip
\smallskip
\smallskip

\noindent \textbf{\large{}Supplementary Notes 6- Repetitive QEC of generalized NP codes.}
\smallskip
\smallskip
\smallskip
\smallskip

The repetitive QEC of bosonic codes have many applications, such as extending the lifetime of qubits or the distance of the quantum communication. In the main text, we have demonstrated the potentiality of generalized NP codes with quantum repeaters to implement the quantum key distribution. Since one round QEC only results in a small bit-flip rate $\epsilon_x$ and phase-flip rate $\epsilon_z$, it can be modeled as a Pauli channel $\mathcal{E}(\hat{\rho})=\mathcal{R}\mathcal{N}(\hat{\rho})=(1-\epsilon_x-\epsilon_z)\hat{\rho}+\epsilon_x\bar{X}\hat{\rho}\bar{X}+\epsilon_z\bar{Z}\hat{\rho}\bar{Z}$. This Pauli channel can also be represented as a scattering matrix
\begin{equation}
    \mathcal{E}=\begin{pmatrix}
    1-\epsilon_x&0&0&\epsilon_x\\
    0&1-\epsilon_x-2\epsilon_z&\epsilon_x&0\\
    0&\epsilon_x&1-\epsilon_x-2\epsilon_z&0\\
    \epsilon_x&0&0&1-\epsilon_x
    \end{pmatrix},
\end{equation}
and $\mathcal{E}[\rho_{0,0},\rho_{0,1},\rho_{1,0},\rho_{11}]^T$ can give the density matrix $\mathcal{E}(\hat{\rho})$ after one round QEC. For $N$ cycles of QEC with $N$ repeater stations, the accumulated bit-flip rate and the phase-flip rate can be obtained as $E_X=\frac{1}{2}(\mathcal{E}^N_{1,4}+\mathcal{E}^N_{4,1})$ and $E_X=\frac{1}{4}[(\mathcal{E}^N_{1,4}+\mathcal{E}^N_{4,1}+\mathcal{E}^N_{1,1}+\mathcal{E}^N_{4,4})-(\mathcal{E}^N_{2,2}+\mathcal{E}^N_{2,3}+\mathcal{E}^N_{3,2}+\mathcal{E}^N_{3,3})]$. Here we consider the quantum key distribution with the four-state protocol~\cite{li2017,Wang2005,Scarani2009},  and the performance of the quantum repeaters can be quantified by the secure key rate per mode (SKRPM)
\begin{align}
   R_{\mathrm{SKRPM}}=1-H(E_x)-H(E_z).
\end{align}
Here $H(p)=-p\mathrm{log}_2p-(1-p)\mathrm{log}_2(1-p)$ is the binary entropy function. Moreover, it is crucial to minimize the error accumulation rate $\tau(\Gamma,\Gamma_{\phi},s,f,\theta_n)=(1-F)/\tilde{L}_0$ for the optimal distance of the quantum communication, where $\tilde{L}_0=L_0/L_{att}$ is the dimensionless repeater spacing. For given optional loss rate $\Gamma$ and dephasing rate $\Gamma_\phi$, it requires to optimize the code parameters ($s,f,\theta_n$) and the repeater spacing $\tilde{L}_0$ for achieving the minimal error accumulation rate and the maximal quantum communication distance. In the main text, we plot the numerically optimal SKRPM for generalized NP codes by selecting $\Gamma=1\%$ and $\Gamma=0.1\%$, and we also provide another situation with $\Gamma=1\%$ and $\Gamma_{\phi}=0.05\%$, where the optional dephasing rate is further suppressed. The detailed simulation data is summarized in Supplementary Table~\ref{table2}.

\begin{table}[thb]
\centering
\caption{Parameters of generalized NP codes for optimized quantum repeaters performance with fiber coupling loss $\epsilon=1\%$.}
\renewcommand{\arraystretch}{1.5}
\begin{tabular}{m{1cm}<{\centering}m{1cm}<{\centering} m{0.5cm}<{\centering} m{0.5cm}<{\centering}m{2cm}<{\centering} m{2cm}<{\centering}m{2cm}<{\centering}m{2cm}<{\centering} m{1.5cm}<{\centering} m{2.5cm}<{\centering}} 
\hline\hline
Lattice&$\Gamma_{\phi}$&$s$&$f$&$\theta_n$&$\epsilon_x$&$\epsilon_z$&$\tau$&$\bar{n}_{\mathrm{code}}$&\begin{tabular}[c]{@{}c@{}}$L_{\mathrm{tot}}~(\mathrm{km})$\\\hline $R=0.1$ $R=0.01$\end{tabular}\\\hline\hline
\multirow{2}{*}{Rect.}&$0.1\%$&$7$&0&$K=15$&$7.9\times10^{-6}$&$7.7\times10^{-6}$&$5.8\times10^{-3}$&$52.5$&535~~~~~~628\\
&$0.05\%$&$10$&$0$&$K=20$&$1.0\times10^{-6}$&$1.0\times10^{-6}$&$1.0\times10^{-3}$&$100$&3102~~~~~~3636\\\hline
\multirow{2}{*}{Obl.}&$0.1\%$&$1$&$7/9$&$r=-1$&$4.7\times10^{-6}$&$6.6\times10^{-6}$&$5.2\times10^{-3}$&$73$&646~~~~~~758\\
&$0.05\%$&$1$&$9/11$&$r=-1$&$5.7\times10^{-7}$&$1.1\times10^{-6}$&$9.3\times10^{-4}$&$105$&3840~~~~~~4512\\\hline
\multirow{2}{*}{Diam.}&$0.1\%$&$4$&1/2&$r=-0.2$&$3.5\times10^{-6}$&$5.6\times10^{-6}$&$3.7\times10^{-3}$&$60$&940~~~~~~1104\\
&$0.05\%$&$5$&$1/2$&$r=-0.2$&$3.3\times10^{-7}$&$8.2\times10^{-7}$&$6.5\times10^{-4}$&$85$&5820~~~~~~6870\\
\hline\hline
\end{tabular}
\label{table2}
\end{table}
\smallskip
\smallskip
\smallskip
\smallskip

\noindent \textbf{\large{}Supplementary Notes 7- Noisy quantum error correction with experimental limitations.}
\smallskip
\smallskip
\smallskip
\smallskip

In Fig. 3 of the main text, we propose using cross-Kerr interaction between the data mode and an ancilla bosonic mode to implement the entangling gate. Such a method is expected to manipulate the NP codes with errors in bosonic level. However, the cross-Kerr interaction strength between two cavities is not large enough compared to the cavity decay rate in the current experiments. Therefore, it represents a long-term goal for the QEC, which only involves bosonic modes.  

For the current experimental consideration, the readout of the number-parity for the NP code can be easily achieved using the traditional dispersively coupled transmon, which is widely employed in bosonic QEC experiments. Furthermore, the teleportation-based QEC between two cavity modes can also be indirectly achieved with an ancilla transmon, as shown in the Eq. (26) of the main text. Based on these considerations, the time-dependent QEC circuit proposed in the main text can be represented in Supplementary Figure~\ref{Fig.S6}. Different from the ideal QEC scenario, the noise from gate operations of finite duration will introduce additional errors.

\begin{figure}[thb]	
\centering
\includegraphics[width=0.9\linewidth]{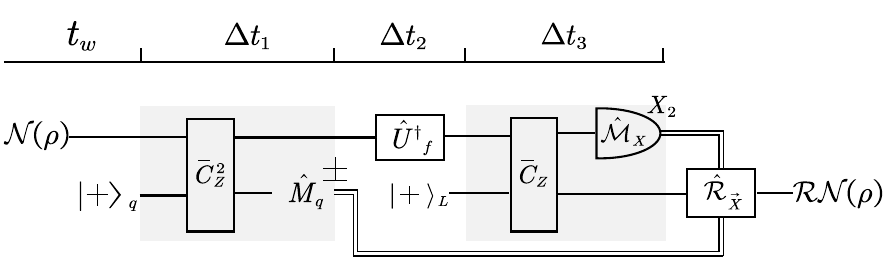}
\caption{The time-dependent version of the teleportation-based QEC scheme for NP codes with current experimental considerations. The number parity readout is realized by a dispersively-coupled discrete-variable ancilla. The direction of the timeline represents the direction of system evolution, where $t_w$ is the waiting time for the bosonic mode to accumulate errors, and $\Delta t_i$ with $i=1,~2,~3$ is the duration of the corresponding gate operations.}
\label{Fig.S6}
\end{figure}

Here we provide a simulation of one round QEC described in Supplementary Figure~\ref{Fig.S6}, with the noise model continuously applied. Then, we compare the QEC performance of the D-NP codes obtained from the time-dependent QEC and the ideal QEC. This numerical simulation is based on the following considerations: (1) The required gate operations have not been simultaneously realized in the same bosonic QEC experiment; thus, we select reasonable parameters according to several experimental works in superconducting quantum circuits. (2) We assume that the state preparation is perfect and focus on the errors from imperfect gates and phase measurements. If the state preparation is faulty, the QEC fidelity cannot exceed that of the state preparation, resulting in a direct, uncorrectable error.

Firstly, the codewords of the D-NP codes used in the QEC cycle are selected as
\begin{equation}
    |\left. + \right> _L=e^{-\frac{\alpha ^2}{2}}\sum_{n=0}^N{\frac{\alpha ^ne^{-i\frac{\pi n^2}{4}}}{\sqrt{n!}}|\left. 2n \right>},\ \ |\left. - \right> _L=e^{-\frac{\alpha ^2}{2}}\sum_{n=0}^N{\frac{\left( -\alpha \right) ^ne^{-i\frac{\pi n^2}{4}}}{\sqrt{n!}}|\left. 2n \right> .}
    \label{eq.s20}
\end{equation}
Here $N$ is the truncation of the Fock space, which we set $N=55$ for the maximal mean photon number $\bar{n}_{\rm{code}}=20$ of the code states. This NP code can be described by the set of parameters $\left[ s=2,\ f=1/2,\ r=0 \right] $ . 
At the beginning of QEC, we assume that the code state preparation is perfect, and the code states take a period of waiting time $t_w$ to accumulate errors. In experiments, the main error of the storage cavity mode is photon loss, while pure dephasing is neglected. Therefore, the Lindblad master equation of the storage cavity mode within the waiting time is 
\begin{equation}
    \frac{\partial\rho}{\partial t}=\gamma\mathcal{L}[\hat{a}]\rho.
\end{equation}
We choose a reasonable decay rate $\gamma=1$ kHz for the cavity mode, which corresponds to a lifetime 1 ms, and an appropriate waiting time $t_w=20~\mu\rm{s}$ ($\gamma t_w=0.02$).

\begin{figure}[thb]	
\centering
\includegraphics[width=0.9\linewidth]{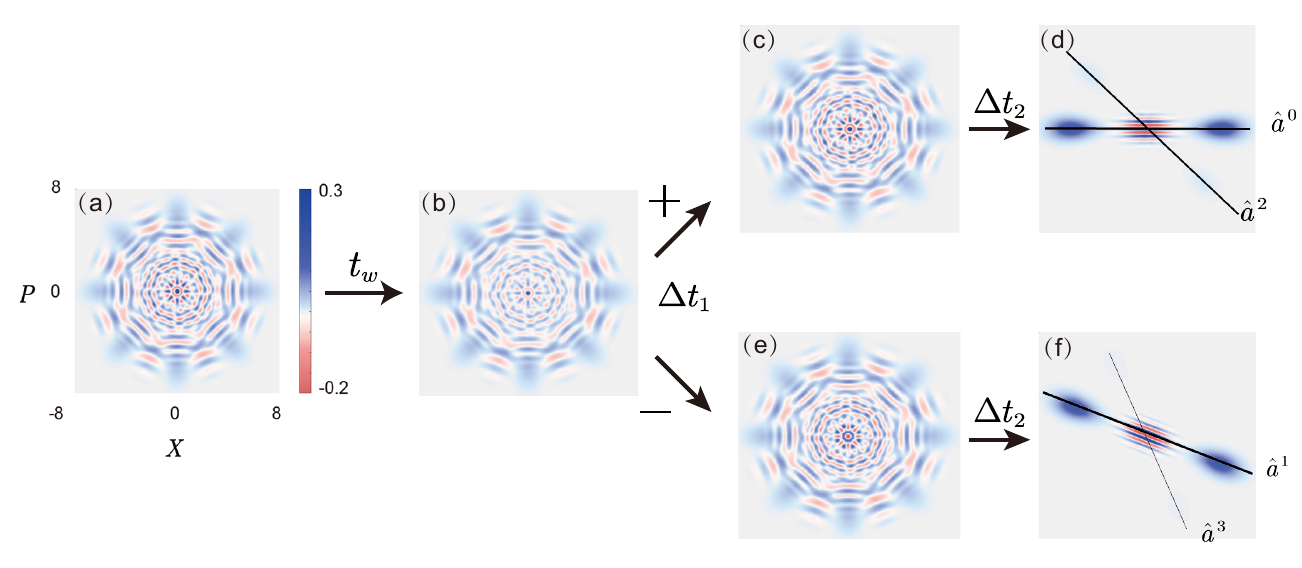}
\caption{The time evolution of the Wigner function for the code state $|+\rangle_L$ in the time-dependent QEC. Here the mean photon number of code states is selected as $\bar{n}_{\rm{code}}=16$ for an example, such a large mean photon number allows the number-phase vortex effect to be obvious. All displayed Wigner functions share the same coordinate range and colorbar. Panels (a) and (b) depict the code state evolution during the waiting time $t_w$, while panels (c) and (d) show the quantum trajectory under even number parity. The complementary trajectory for odd parity is displayed in panels (e) and (f). }
\label{Fig.S7}
\end{figure}

Then, we simulate the number parity readout of the cavity mode, which is the first step of QEC. It is based on the entangling gate $\bar{C}_q$ between the cavity and a ancilla transmon, which is based on the dispersive coupling interaction $H_{\Delta t_1}=\chi\hat{n}\otimes|e\rangle\langle e|$, with a reasonable interaction strength $\chi=2\pi\times 4$ MHz, and the gate time $\Delta t_1=\pi/\chi$. During the gate time, the evolution of the two-body system is 
\begin{equation}
    \frac{\partial \rho }{\partial t}=-i\left[ H_{\Delta t_1},\rho \right] +\gamma \mathcal{L}\left[ \hat{a} \right] \rho+\kappa \mathcal{L}\left[ \sigma _- \right] \rho .
\end{equation}
The first term on the right side of the equation is the unitary evolution with the dispersive coupling interaction. The second and third terms represent the energy decay of the cavity and the transmon, respectively. We choose the transmon decay rate $\kappa=10$ kHz, and the transmon relaxation will introduce dephasing errors to the cavity modes. These parameters of the transmon are selected from Ref.~\cite{Ni2023}, where the pure dephasing of the transmon is negligible. At the end of the gate time, the cavity and the transmon are entangled, and the measurement outcomes “+” and “-” of the transmon indicate the even and odd parity of the cavity mode. 

The next step is to implement the interface gate, which is based on the tunable self-Kerr interaction $H_{\Delta t_2}=\frac{K}{2}\hat{n}^2$ of the cavity mode. Ref.~\cite{he2023fast} reported the tunable range of the Kerr nonlinearity of cavity mode from  $-2\pi\times 5$  MHz to $2\pi\times 6$ MHz. Here we select  MHz for a reasonable consideration. The gate time of the interface gate is  , and the D-NP codes will be transmitted to the R-NP codes at the end of the gate time. The evolution of the cavity mode is
\begin{equation}
    \frac{\partial \rho}{\partial t}=-i\left[ H_{\Delta t_2},\rho \right] +\gamma \mathcal{L}\left[ \hat{a} \right] \rho. 
\end{equation}

We show the time evolution of the Wigner function for the code state $|+\rangle_L $ in Supplementary Figure~\ref{Fig.S7}. The number parity measurement can only distinguish the 0-th order and the first-order photon loss error, while the higher-order photon loss errors can be identified via the phase rotation syndromes induced by number-phase vortex effect. Although photon loss in the Kerr medium introduces dephasing errors to the bosonic mode, it does not eliminate the number-phase vortex effect; instead, it increases the phase uncertainty of the codewords.

\begin{figure}[thb]	
\centering
\includegraphics[width=0.9\linewidth]{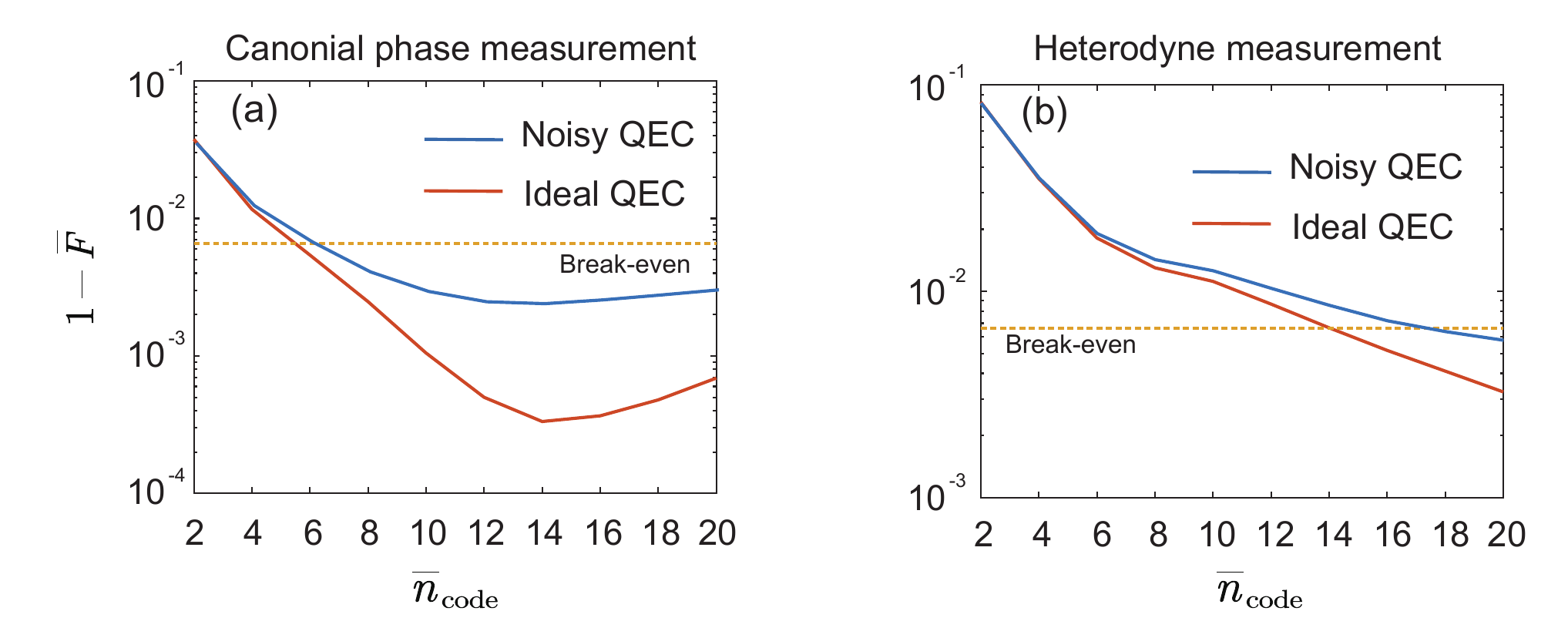}
\caption{The performance of noisy and ideal QEC for the D-NP codes (Eq.~\ref{eq.s20}) with different mean photon number. The QEC performance is quantified by the average fidelity between the recovered code states and the corresponding initial code states. The break-even line is estimated by the Fock states $|0\rangle$ and $|1\rangle$ coding. Results (a) and (b) are obtained with the canonical phase measurement and heterodyne measurement, respectively.    }
\label{Fig.S8}
\end{figure}
The last step is to teleport the quantum state from the noisy cavity mode to another fresh cavity mode. This teleportation can be implemented via a cascaded teleportation with an ancilla transmon, as shown in Eq.~(26) of the main text. The entangling gate Hamiltonian of the cavity mode and the transmon is also a dispersive interaction $H_{\Delta t_3}=\chi\hat{n}\otimes|e\rangle\langle e|$. The time evolution of the three-body system during the implementing controlled-phase gate between the cavity mode and the transmon is 
\begin{equation}
    \frac{\partial \rho}{\partial t}=-i\left[ H_{\Delta t_3}^{j},\rho  \right] +\gamma \mathcal{L}\left[ \hat{a}_1 \right] \rho +\kappa \mathcal{L}\left[ \sigma _- \right] \rho +\gamma \mathcal{L}\left[ \hat{a}_2 \right] \rho .
\end{equation}
Here $H_{\Delta t_3}^{j}$ with $j=1,~2$ corresponds to dispersive interaction between the transmon and j-th bosonic mode with the annihilation operator $\hat{a}_j$. The successive controlled-phase gates will take the total gate duration $\Delta t_3=\pi/\chi$. 

Lastly, a phase measurement and a qubit measurement are performed on the cavity mode and the ancilla transmon, respectively. It teleports the quantum state from the noisy cavity mode to a fresh cavity mode. Here we consider two different phase measurements, the canonical phase measurement and the heterodyne measurement, their POVM elements can be written as
\begin{equation}
    \hat{\mathcal{M}}_{x}^{\rm{Can}}=\frac{1}{N}\sum_{n=0}^N{e^{i\left( m-n \right) x}|\left. m \right> \left< n| \right.},~~
    \hat{\mathcal{M}}_{x}^{\rm{Het}}=\sum_{n=0}^N{\frac{\Gamma \left[ \left( n+m \right) /2+1 \right]}{\sqrt{\Gamma \left( n+1 \right) \Gamma \left( m+1 \right)}}e^{i\left( m-n \right) x}|\left. m \right> \left< n| \right.}
\end{equation}
The heterodyne measurement is an approximation of the canonical phase measurement with a lower efficiency, but its implementation is more feasible compared to the canonical phase measurement. We simulate the noisy and ideal QEC performance of the D-NP code with the two different phase measurements. The recovery feedback does not need to be explicitly applied to every QEC cycle for quantum memory; instead, logical operations are recorded based on each measurement outcome and recovered after decoding from logical to physical qubits.

We compare the noisy and ideal QEC performance of the D-NP code, as shown in Supplementary Figure~\ref{Fig.S8}. Here, the average infidelity of ideal QEC reflects uncorrectable errors in the teleportation-based QEC scheme, while the average infidelity difference between ideal and noisy QEC reveals noise induced by imperfect gate operations. This discrepancy is particularly pronounced at an infidelity of noisy QEC near $1\times 10^{-3}$, as operational errors dominate in this regime. Our analysis shows that operational errors primarily stem from transmon dissipation during entangling gates, which can be estimated as
\begin{equation}
    \kappa \left( \Delta t_1+\Delta t_3 \right) =\frac{2\pi \kappa}{\chi}=2.5\times 10^{-3}.
\end{equation}
The uncorrectable errors hinder the attainment of higher QEC performance. As discussed in the section "Experimental Feasibility and Limitations," a straightforward approach to reducing uncorrectable errors is to encode the transmon in a higher excited state for achieving the second-order error $ (\kappa \Delta t)^2$ of a transmon. Moreover, developing a fault-tolerant QEC scheme for bosonic codes represents the fundamental solution to mitigate operational errors.

\bibliography{reference_NPcode.bib}